\documentclass[12pt]{article}
\usepackage{a4,epsfig}
\usepackage{graphicx,longtable}
\usepackage{color,amssymb}
\usepackage{here}
\newcommand{\ba}{\begin{eqnarray}}
\newcommand{\ea}{\end{eqnarray}}
\newcommand{\ban}{\begin{eqnarray*}}
\newcommand{\ean}{\end{eqnarray*}}
\newcommand{\be}{\begin{equation}}
\newcommand{\ee}{\end{equation}}

\begin{document}
\vspace*{-0.5in}

\begin{flushright}

LPT Orsay 10-13 \\

\end{flushright}

\vskip 5pt

\begin{center}
{\Large{\bf One-loop correction effects on supernova neutrino fluxes:
\\ \ a new possible probe for Beyond Standard Models. }}\\

\vspace{1cm}
{\large J. Gava}\footnote{gava@ipno.in2p3.fr},
{\large C.-C. Jean-Louis}\footnote{charles.jean-louis@th.u-psud.fr}

$^1$ Institut de Physique Nucl\'eaire, B\^at. 100, CNRS/IN2P3 UMR
8608 et Universit\'e Paris-Sud 11, F-91406 Orsay cedex, France

$^2$ Laboratoire de Physique Th\'eorique, B\^at. 210, CNRS UMR
8627 et Universit\'e Paris-Sud 11, 91405 Orsay Cedex, France

\end{center}
\vspace{2.cm}
\begin{abstract}
\noindent We present the consequences of a large radiative
correction term coming from Supersymmetry (SUSY) upon the electron neutrino fluxes
streaming off a core-collapse supernova using a 3-flavour
neutrino-neutrino interaction code. We explore the interplay
between the neutrino-neutrino interaction and the effects of the
resonance associated with the $\mu-\tau$ neutrino index of
refraction. We find that sizeable effects may be visible in the
flux on Earth and, consequently, on the number of events upon the
energy signal of electron neutrinos in a liquid argon detector.
Such effects could lead to a probe for Beyond Standard Model (BSM) physics and, ideally, to
constraints in the SUSY parameter space.
\end{abstract}

\date{\today}
\newpage
\pagestyle{plain}
\setcounter{page}{1}
\setcounter{footnote}{0}


\section{Introduction}

During the past decade, our comprehension of the neutrino
interactions in a supernova environment has deeply evolved, due
 to the increasing precision of neutrino experiments and
computational force. In addition with the traditional MSW effect
~\cite{Wolfenstein:1977ue,Mikheev:1986wj}, the neutrino-neutrino
interaction has been proven to be fundamental for the neutrino
evolution in such environment. This interaction, whose effects
have changed the existing paradigm of SN neutrino physics, has
been recently the subject of intense investigation
\cite{Pantaleone:1992eq,Sigl:1992fn,Samuel:1993uw,Qian:1994wh,Sigl:1994hc,
Pastor:2001iu,Pastor:2002we,Balantekin:2004ug,Duan:2005cp,Duan:2006an,
Hannestad:2006nj,Balantekin:2006tg,Duan:2006jv,Fogli:2007bk,Raffelt:2007cb,
Raffelt:2007xt,Duan:2007sh,Dasgupta:2007ws,EstebanPretel:2008ni}.
For a recent extensive review on this subject, see
\cite{Duan:2009cd}.

In addition with these interactions, an other matter interaction $V_{\mu \tau}$, which arises
at one-loop order ($O\left({\alpha\over \pi\sin^2\theta_W}{m_\tau^2\over M_W^2}\right)$) makes
 the mu and tau neutrino index of refraction different \cite{Botella:1986wy,Roulet:1995ef}.
This radiative correction potential is related to the
charged-current matter potential $V_c$, in an electrically neutral
medium, by a parameter $\varepsilon$, which is defined as
$-p_\nu \Delta n_{\tau\mu}=V_{\mu\tau}=\varepsilon V_e$\footnote{We do not consider the presence of other charged leptons than electrons in the supernova environment. Therefore, $V_c=V_e$.}. The value of $\varepsilon$ is $5.4\times\,10^{-5}$ in the Standard Model. Only a few
papers have shown the interest of such radiative correction term.
Without the neutrino-neutrino interaction, $V_{\mu\tau}$ has been
shown to present effects on Earth if the $\nu_\mu$ and $\nu_\tau$
fluxes are different at emission \cite{Akhmedov:2002zj}. Via a
non-zero CP-violating phase, $V_{\mu\tau}$ can also in theory
induce effects on the (anti-) electron neutrino fluxes in the Sun
\cite{Minakata:1999ze} or in a supernova environment
\cite{Balantekin:2007es,Kneller:2009vd}. Including the collective
effects, the CP-violation effects due to $V_{\mu \tau}$ turn out
to be larger \cite{Gava:2008rp}, and the electron neutrino flux
displays a $\theta_{23}$ dependence in presence of a large
$V_{\mu\tau}$ coming from a high density profile
\cite{EstebanPretel:2007yq,Duan:2008za}.

Despite its success, the Standard Model (SM) of particle physics
exhibits a number of shortcomings which might be remedied by new
Physics showing up at the Terascale. One very popular extension of
the SM is Supersymmetry (SUSY)~\cite{WessBagger}. Virtues of SUSY are numerous: it
can provide a natural candidate to explain dark matter and it allows the elimination of the hierarchy problem and the unification of the gauge couplings at the scale of
grand unification ($M_{GUT}$).

In a previous paper, we have calculated all one-loop radiative
correction terms to the matter interaction in the SUSY
framework~\cite{Gava:2009gt}. SUSY particles like sleptons,
squarks up and down, charged Higgs, neutralinos and charginos take
place in these loops. With a dedicated numerical routine, we have
scanned and identified regions in the parameter space that yield
interesting value of the $V_{\mu \tau}$ potential, namely large
values up to $\varepsilon_{SUSY}\simeq 2.10^{-2}$.

The recent simulations of neutrino evolution
\cite{Gava:2009pj,Galais:2009wi} in supernova take into account a
dynamical density profile, in addition with the neutrino
collective effects. All interactions combined show a
characteristic imprint in the anti-neutrino fluxes, whose
detection could yield precious information concerning the dynamics
of the density profile or fundamental neutrinos properties like
the hierarchy or the value of the third mixing angle
$\theta_{13}$. With larger value of $\varepsilon$ than in the SM
case, we show that $V_{\mu \tau}$ can exhibit sizeable and
characteristic effects on the electron neutrino fluxes
particularly in the inverse hierarchy where the $\nu-\nu$
interaction display salient features such as the synchronization
regime, the bipolar transition and the spectral split. Such BSM
imprints could be seen in a neutrino observatory on Earth and
could ideally lead to constraints in the SUSY parameter space.

The paper is organized as follows. In Sec.~\ref{sec:theoframe} we
explicit the evolution equations which lead to the calculation of
the electron neutrino flux, describe the framework used, and
justify the approximations taken to yield our results. In
Sec.~\ref{sec:typical} we describe the three typical behaviours
that can occur due to the interplay between the neutrino-neutrino
interaction and the $\mu-\tau$ resonance. In
Sec.~\ref{sec:effectepsilon} and Sec.~\ref{sec:parameters}, we
analyze the role of the parameters that can influence such
interplay and consequently the $\nu_e$ flux as a function of
energy.
The subsections \ref{sec:effectlum} and \ref{sec:effectdens} are
dedicated to the study of the consequence of varying the
luminosity and the density which is equivalent to looking at the flux at
different times. We then use these fluxes to display the number
of events in a liquid argon detector that we describe in
Sec.~\ref{sec:detectEarth}. Finally, before concluding in
Sec.~\ref{sec:conclusion}, we discuss in
Sec.~\ref{sec:discussion}, about the possibility of the
observation of such beyond standard effects for realistic
conditions.

\section{Theoretical framework}
\label{sec:theoframe}

The main goal of this paper is to explore the impacts of the
parameters that influence the interplay between the
neutrino-neutrino interaction and the $\mu-\tau$ resonance and
consequently study the possibility that such effects can be
observable in a neutrino observatory in order to get a possible
probe for models BSM. In this section, we first present the
evolution equation we have numerically solved and the hypothesis
used to yield the results of the following sections.

\subsection{Equation of evolution.}
\label{sec:Eqevolution}

\noindent In a dense environment the non-linear coupled neutrino
evolution  equations with neutrino self-interactions are given
by\footnote{The dependence on r is equivalent to the dependence in
t since we suppose that neutrinos travel at light speed in the
supernova and we take $c=1$.}:
\begin{equation}
\label{e:1}
i{ d \over{dt}} \psi_{{\nu}_{\underline{\alpha}}} =[H_0 + H_m + H_{\nu \nu}] \psi_{{\nu}_{\underline{\alpha}}}
\end{equation}
where $ \psi_{{\nu}_{\underline{\alpha}}}$ denotes a neutrino
created at the neutrinosphere initially in a flavour state $\alpha
=e, \mu,\tau$, $H_0 = U H_{vac} U^{\dagger}$ is the Hamiltonian
describing the vacuum oscillations $H_{vac}=\mathrm{diag}(E_1,E_2,E_3)$,
$E_{i=1,2,3}$ being the energies of the neutrino mass eigenstates,
and $U$ the unitary Maki-Nakagawa-Sakata-Pontecorvo matrix
\begin{equation}
\label{e:2}
U = T_{23} T_{13} T_{12} = \left(\matrix{
     1 & 0 & 0  \cr
     0 &  c_{23}  & s_{23} \cr
     0 & - s_{23} &  c_{23} }\right)
 \left(\matrix{
     c_{13} & 0 &  s_{13} e^{-i\delta}\cr
     0 &  1 & 0 \cr
     - s_{13} e^{i\delta} & 0&  c_{13} }\right)
 \left(\matrix{
     c_{12} & s_{12} &0 \cr
     - s_{12} & c_{12} & 0 \cr
     0 & 0&  1 }\right) ,
\end{equation}
$c_{ij} = \cos \theta_{ij}$ ($s_{ij} = \sin \theta_{ij}$) with
$\theta_{12},\theta_{23}$ and $\theta_{13}$ the three neutrino
mixing angles. The presence of a Dirac $\delta $ phase in
Eq.(\ref{e:2}) renders $U$ complex and introduces a difference
between matter and anti-matter.

The neutrino interaction with matter is taken into account through
an effective Hamiltonian which corresponds to the diagonal matrix
$H_m=diag(V_e,0 ,V_{\mu \tau})$. $V_e (r) = \sqrt{2} G_F n_e
(r)$ is the matter potential due to the charged-current
interaction between electron (anti-) neutrinos and the electrons
present in the medium where $G_F$ is the Fermi coupling constant
and $n_e (r)$ is the electron density in the star. At
the tree level, neutral current interactions introduce an overall
phase only. Note that this density will imply in supernova possibly two resonances when the MSW condition \cite{Mikheev:1986wj} (here shown for two flavours) is reached:
\be
\frac{\Delta m^2}{2E_{\nu}}\cos2\theta=V_e(r)
\ee
Depending on the sign of $\Delta m_{32}^2$ ,the "H-resonance" (high density) may happen either for neutrinos or anti-neutrinos while the "L-resonance" (low density) happens for neutrinos only and is dependent on $\Delta m_{21}^2$.

As previously mentioned, in the Standard Model case, Botella et
al. in \cite{Botella:1986wy} have highlighted the presence of a
one-loop matter potential arising from radiative corrections to
neutral-current $\nu_\mu$ and $\nu_\tau$ scattering. Such matter
potential can therefore be seen as an effective presence of $\tau$
particles and one can write \be \label{e:vmutau} V_{\mu\tau}=
\sqrt{2}\,G_{\rm F}Y_{\tau, SM}^{\rm eff}n_B \ee
 where $n_B$ is the baryon density inside the supernova and
\begin{equation}
 Y_{\tau, SM}^{\rm eff}=\frac{3\sqrt{2}\,G_{\rm F}m_\tau^2}{(2\pi)^2}
 \left[\ln\left(\frac{m_W^2}{m_\tau^2}\right)-1+\frac{Y_n}{3}\right]
 =2.7\times10^{-5}\,,
\label{eq:Ytau}
\end{equation}
The value of $Y_{\tau, SM}^{\rm eff}$ is obtained using the hypothesis of a isoscalar
medium i.e, $n_e=n_p=n_n$ ($n_p$ and $n_n$ being respectively the proton and neutron density in the star), therefore the neutron fraction $Y_n=n_n/(n_p+n_n)$ is equal to 0.5. For our convenience, we have
defined the overall radiative correction factor as : \be
\label{e:epsilon} \varepsilon=\frac{Y_{\tau, SM}^{\rm
eff}+Y_{\tau, SUSY}^{\rm eff}}{Y_e} \ee where $Y_{\tau, SUSY}^{\rm
eff}=0$ in the Standard Model and the electron fraction $Y_e=n_e/(n_n+n_p)$ is taken to be 0.5 all
along the paper.

The general form of the neutrino self-interaction term is
\begin{equation}
 \label{e:4}
H_{\nu \nu} = \sqrt{2} G_F
\sum_{\alpha} \sum_{\nu_{\alpha},\bar{\nu}_{\alpha}} \int
 \rho_{{\nu}_{\underline{\alpha}}} ({\bf q}')(1 - {\bf \hat{q}} \cdot {\bf \hat{q}'}) dn_{\alpha} dq'
\end{equation}
where
$\rho = \rho_{{\nu}_{\underline{\alpha}}}$  ($- \rho^*_{{\nu}_{\underline{\alpha}}}$) is
the density matrix for neutrinos (antineutrinos)
\begin{equation}\label{e:5}
 \rho_{{\nu}_{\underline{\alpha}}} = \left(\matrix{
     |\psi_{\nu_e}|^2 & \psi_{\nu_e} \psi_{\nu_{\mu}}^*&   \psi_{\nu_e} \psi_{\nu_{\tau}}^* \cr
     \psi_{\nu_e}^* \psi_{\nu_{\mu}}  &   |\psi_{\nu_{\mu}}|^2     &   \psi_{\nu_{\mu}} \psi_{\nu_{\tau}}^* \cr
      \psi_{\nu_e}^* \psi_{\nu_{\tau}} &   \psi_{\nu_{\mu}}^* \psi_{\nu_{\tau}} &   |\psi_{\nu_{\tau}}|^2   }\right) ,
\end{equation}
${\bf q}$ (${\bf q'}$) denotes the momentum of the neutrino of
interest (background neutrino) and $dn_{\alpha}$ is the
differential number density. The emission geometry is based on the
so called ``bulb model'' with spherical symmetry \cite{Duan:2006an}: neutrinos are
assumed to be half-isotropically emitted from the neutrinosphere.

In the multi-angle treatment, SN neutrinos travel on different trajectories. This is taken into account by
the factor $(1 - {\bf \hat{q}} \cdot {\bf \hat{q}'})$ in
Eq.(\ref{e:4}). The single angle approximation consists of
averaging such factor along the polar axis, i.e. $ \rho ({\bf q}) = \rho (q)$, Eq.(\ref{e:4}) reduces to
\begin{equation}\label{e:4bis}
H_{\nu \nu} ={ \sqrt{2} G_F \over {2 \pi R_{\nu}^2}} D(r/ R_{\nu}) \sum_{\alpha} \int [\rho_{{\nu}_{\underline{\alpha}}} (q') L_{{\nu}_{\underline{\alpha}}} (q') - \rho_{\bar{{\nu}}_{\underline{\alpha}}}^*(q')  L_{\bar{{\nu}}_{\underline{\alpha}}}(q')] dq'
\end{equation}
with the geometrical factor
\begin{equation}\label{e:4tris}
D(r/R) = {1 \over 2} \left[1 - \sqrt{1 - \left({R \over r}\right)^2 }\right]^2
\end{equation}
where the radius of the neutrino sphere is $R = 10$km. In a spherically symmetric environment, a single-angle treatment tends to capture the main non-linear behaviour.

For this paper, we shall adopt
\begin{equation}\label{e:4quadris}
L_{{\nu}_{\underline{\alpha}}}(E_{\nu})= {L^0_{{\nu}_{\underline{\alpha}}} \over{T_{\nu_{\underline{\alpha}}}^3 \langle E_{\nu_{\underline{\alpha}}} \rangle F_2(\eta)}}{{E_{\nu_{\underline{\alpha}}}^2} \over{1 + \exp{(E_{\nu_{\underline{\alpha}}}/T_{\nu_{\underline{\alpha}}} - \eta)}}}
\end{equation}
as the neutrino luminosity, where $F_2(\eta)$ is the Fermi integral,
$L^0_{{\nu}_{\underline{\alpha}}}$ and
$T_{\nu_{\underline{\alpha}}}$ are the luminosity and temperature
at the neutrinosphere. $\langle E_{\nu_{\underline{\alpha}}}
\rangle$ is the average neutrino energy of the corresponding
flavour $\alpha$.

As we are going to see in the following sections, the neutrino-neutrino interaction is characterized by three distinct phases for the neutrino evolution: the synchronization regime, the bipolar transition and the spectral split. The synchronized regime takes place in the first 50 km outside the neutrino sphere. In this phase, the strong neutrino-neutrino interaction makes neutrinos of all energies oscillate with the same frequency so that flavour conversion is frozen \cite{Pastor:2001iu,Duan:2006jv}.
 When the neutrino self-interaction term diminishes, large bipolar oscillations appear that produce strong flavour conversion for both neutrinos and anti-neutrinos for the case of inverted hierarchy, as long as the $\theta_{13}$ value \cite{Hannestad:2006nj} is non-zero. Eventually neutrinos show complete (no) flavour conversion for energies larger (smaller) than a characteristic energy $E_c \simeq 8$MeV which can be related to the lepton number conservation \cite{Raffelt:2007cb}. This is the spectral split phenomenon: while electron neutrinos swap their spectra with muon and tau neutrinos; the electron anti-neutrinos show a complete spectral swapping. Such behaviours are found for both large and small values of the third neutrino mixing angle, contrary to the standard MSW effect.

We have chosen here to describe the evolution equation using wave-functions such
as in \cite{Duan:2006an,Gava:2008rp} for instance. Note that
equivalently the density matrix or the polarization vector
formalism can be used to describe the neutrino evolution in the
supernova environment.
\subsection{Work hypothesis.}
\label{sec:workhypothesis}

We give the general framework and the hypothesis that we make to obtain our numerical results.

The results we obtained use the best fit
oscillation parameters to date, i.e. $\Delta m^2_{21}= 7.6 \times
10^{-5}$eV$^2$, sin$^2 2\theta_{12}=0.87$ and $\mid\Delta
m^2_{32}\mid=\mid \Delta m^2_{atm}\mid= 2.43 \times 10^{-3}$eV$^2$ for the solar and
atmospheric differences of the mass squares\footnote{Note that
$E_i-E_j \simeq \frac{m_i^2-m_j^2}{2E} = \frac{\Delta
m^2_{ij}}{2E}$} and mixings, respectively \cite{Amsler:2008zzb}.
Apart from Sec.~\ref{sec:effecttheta}, we choose
$\theta_{23}=40^{\circ}$ in the first octant, i.e equivalent to
$\sin^2 2\theta_{23}=0.97$. Unless notified, the Dirac phase is
taken to be zero and $\sin^2 2\theta_{13}=0.1$. This value of $\theta_{13}$ taken is in agreement with the current experimental upper-limit \cite{Amsler:2008zzb}.

During a supernova explosion, three distinct phases take place:
the prompt deleptonization burst, the accretion phase, and the
cooling phase. In this paper, we will focus on the neutrino signal
produced during the early cooling phase where the density of
neutrinos is sufficiently important to induce sizeable non-linear
behaviour. Concerning the cooling phase, the hypothesis of equal luminosities and
equipartition of energies among all neutrino flavours at the
neutrino-sphere is made for the neutrino signal upon which we are
focusing. We also make the assumption of an exponential
decrease of the luminosity \be L_{\nu}=L_{\nu,0} \times
\exp(-t/\tau) \label{e:lumexp} \ee

with $L_{\nu_0}=10^{52}$ erg $\cdot$ s$^{-1}$ and $\tau=3.5$ s the
typical cooling time. Apart from Sec.~\ref{sec:effectlum} and
Sec.~\ref{sec:detectEarth}, we look at time $t\simeq 8$s after
post-bounce which corresponds to  $L_{\nu}= 10^{51}$ erg
$\cdot$ s$^{-1}$ . We consider the hierarchy i.e. $ \langle E_{\nu_e}
\rangle < \langle E_{\bar{\nu}_e} \rangle < \langle E_{\nu_{x}}
\rangle$ with typical values of 10, 15 and 24 MeV respectively. These values correspond approximatively to the cooling case I in \cite{GilBotella:2003sz}.

In this paper, one of the observables is the $\nu_e$ flux, which can be written \ba
\Phi_{\nu_{e}} &=& P(\nu_e\rightarrow \nu_e)F^R_{\nu_{e}}+ P(\nu_e\rightarrow \nu_{\mu})F^R_{\nu_{\mu}}
+P(\nu_e\rightarrow \nu_{\tau})F^R_{\nu_{\tau}} \nonumber\\
&=& P(\nu_e\rightarrow \nu_e)(F^R_{\nu_{e}}-F^R_{\nu_{\mu}})+F^R_{\nu_{\mu}}
\label{e:nueflux} \ea using the unitarity of the evolution
operator of Eq.(\ref{e:1}) and where $F^R_{\nu_{\alpha}}$ is the
flux of flavour $\alpha$ emitted initially. We here made the hypothesis of equal $\nu_\mu$ and $\nu_\tau$ initial fluxes. We can link it to the luminosity $L_{{\nu}_{\underline{\alpha}}}(E_{\nu})$
 defined in Eq.(\ref{e:4quadris}) via the relation:

\be
F^R_{\nu_{\alpha}}= \frac{1}{4\pi R^2}L_{{\nu}_{\underline{\alpha}}}(E_{\nu})
\label{e:link}
\ee
Moreover, we define the flux received on Earth by
\be
\Phi^{Earth}_{\nu_{e}}=<\Phi_{\nu_{e}}>= <P(\nu_e\rightarrow \nu_e)>_{\rm{exit}}(F^R_{\nu_{e}}-F^R_{\nu_{\mu}})+F^R_{\nu_{\mu}}
\ee
The probability exiting the SN $<P(\nu_e\rightarrow \nu_e)>_{\rm{exit}}$, and therefore the flux on Earth, is averaged over distance because of the decoherence phenomenon that takes place for neutrinos travelling over large distance in vacuum \cite{Dighe:1999bi}.
As the neutrinos stream off the supernova, a shock wave propagates
through the SN envelope coming with a reverse shock. This induces
a dynamical density profile which leads to the breaking of the
adiabaticity of the H-resonance and possibly to a multiple
resonance effect. Such interplay with the neutrino fluxes may
imply a salient fingerprint on the signal received in an
observatory on Earth as pointed out in
\cite{Tomas:2004gr,Fogli:2004ff,Kneller:2007kg,Gava:2009pj}.

All along the paper we will focus on the electron neutrino flux in
the inverted hierarchy. Consequently, the shock-wave effect should
not impact on the electron neutrino flux since  no H-resonance
occurs. Therefore, for our electron number density, we adopt the
usual simple analytical profile for the :
\begin{equation}
\label{eq:lambda(r)}
  \lambda(r) =\sqrt{2}G_Fn_B=
    \lambda_r\,\left(\frac{R}{r}\right)^3 \,.
\end{equation}
 where R is equal to $10$km.
Apart from Sec.~\ref{sec:effectdens} and
Sec.~\ref{sec:detectEarth}, we take for reference $\lambda_r=5.7 \times
10^5$km$^{-1}$ which corresponds to a late time density profile.

 In~\cite{Mirizzi:2009td}, SM one-loop corrections to $\nu-\nu$ interaction have been studied
and turned out to be of the same order as $Y_{\tau, SM}^{\rm eff}$
or even smaller. Though they have not been calculated
yet\footnote{A first related study can been found in
\cite{Blennow:2008er}.}, supersymmetric one-loop corrections to
$\nu-\nu$ interaction will be similar to the SM ones but only
larger in the same way than for $Y_{\tau, SUSY}^{\rm eff}$ in
comparison with $Y_{\tau, SM}^{\rm eff}$. For $\varepsilon
\lesssim 2.10^{-2}$, we expect them to be negligible. All the hypothesis made in this
section will be discussed in Sec. \ref{sec:discussion}.

\section{Typical behaviour of a large $\mu$-$\tau$ radiative correction}
\label{sec:typical}

In this section we review the consequences of a large $V_{\mu
\tau}$ on the $\nu_e$ flux in the inverted hierarchy, in presence
of the neutrino-neutrino interaction. Unlike in
\cite{EstebanPretel:2007yq}, we do not increase the value of the
density $\lambda_r$ to obtain a large effective $\tau$ matter
potential but we use instead the radiative correction term
$\varepsilon$ whose value can rise depending on the SUSY
parameters \cite{Gava:2009gt}. Consequently, varying the
$\varepsilon$ for a given energy will only move the $\mu-\tau$
resonance above the synchronized region leaving unchanged the
position of the L-resonance.

\begin{figure}[h]
\vspace{.6cm}
\centerline{\includegraphics[scale=0.25,angle=0]{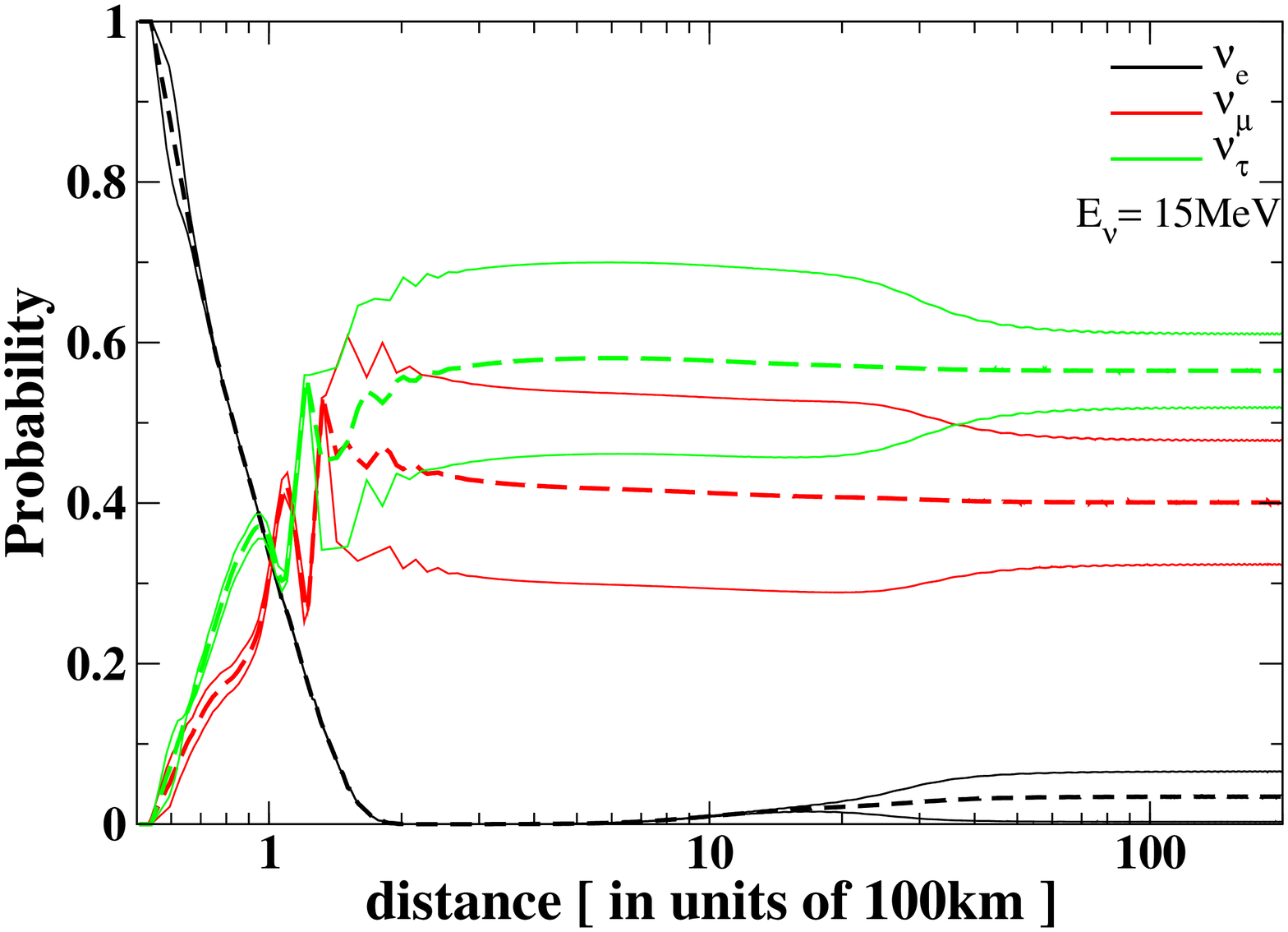}\hspace{.1cm}
\includegraphics[scale=0.25,angle=0]{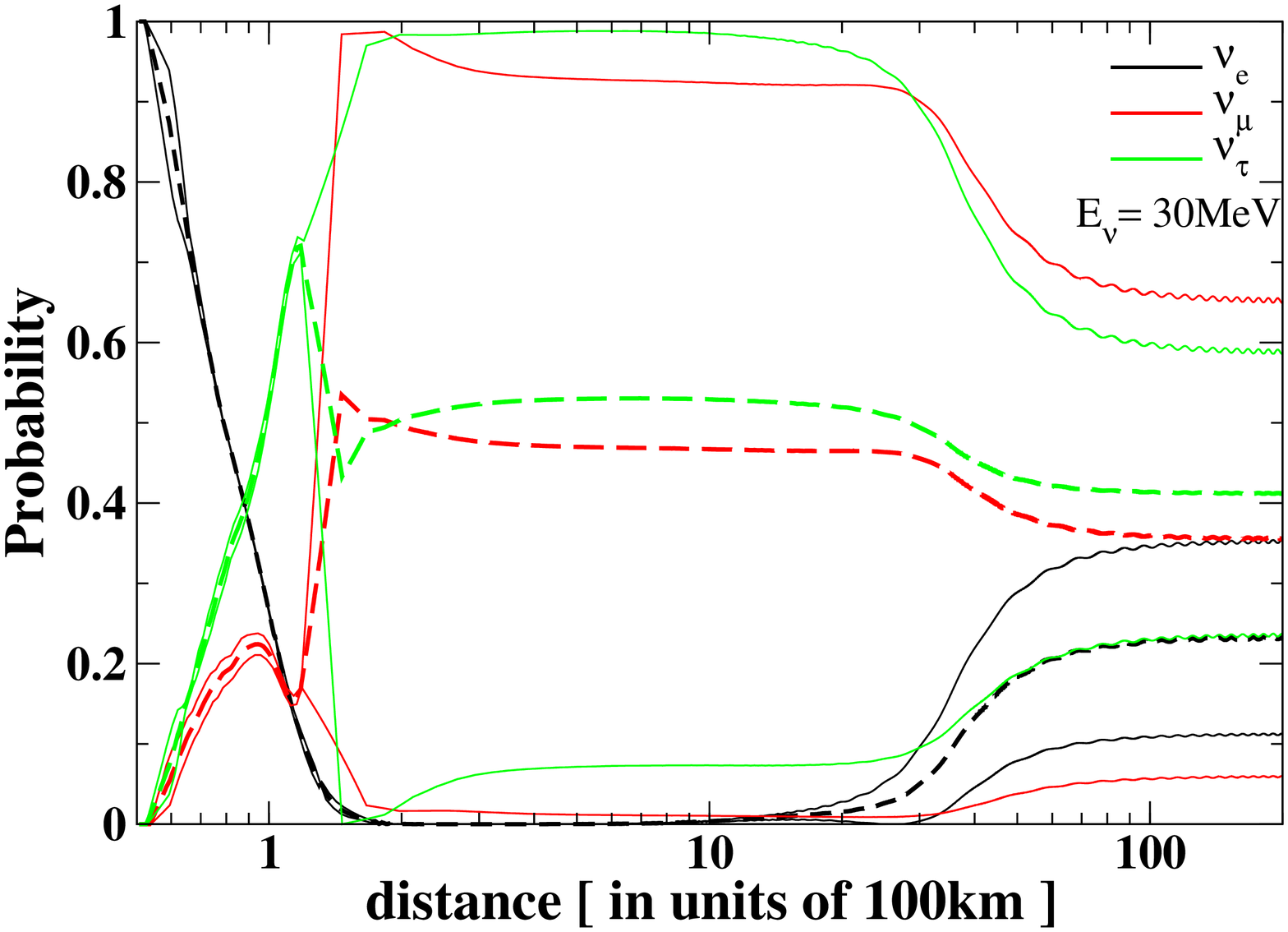}\hspace{.1cm}
\includegraphics[scale=0.25,angle=0]{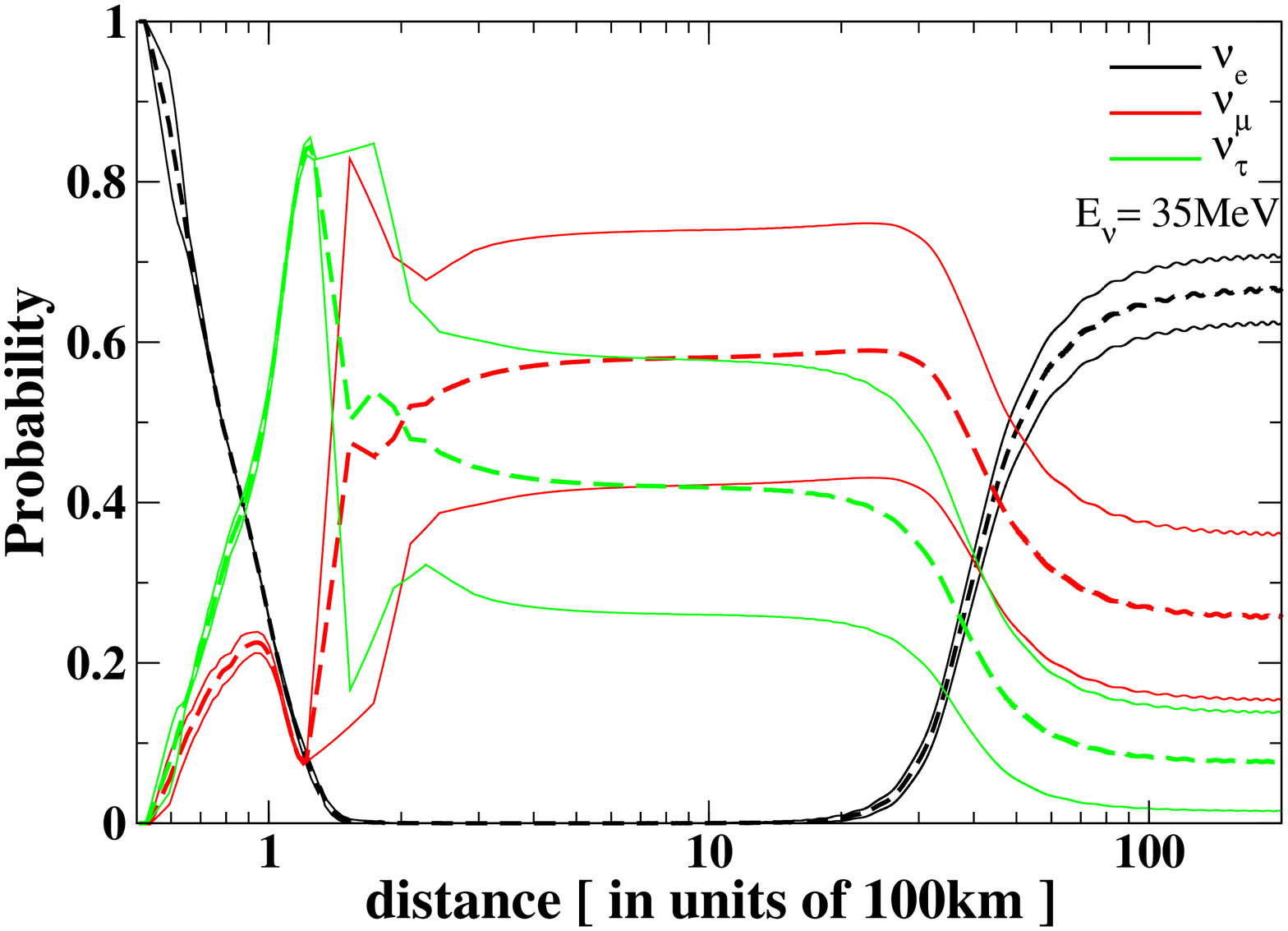}}
\caption{Oscillation probabilities for a $\nu_e$ initially emitted as a function of the distance
from the neutrinosphere. We show the average as dashed lines and the envelopes of the fast-oscillating curves as solid lines. The left, middle and right figures
correspond respectively to type A, type B and type C behaviour with
respective neutrino energy of 15, 30 and 35 MeV.
The value of the radiative correction term is $\varepsilon_r =
10^{-3}$.}
\label{fig:probaeps10-3}
\end{figure}

In Fig.(\ref{fig:probaeps10-3}) we show the oscillation
probabilities for three different energies $E_\nu = 15$, $30$ and $35$
MeV with $\varepsilon=\varepsilon_r = 10^{-3}$ in order to identify precisely
the physical effects appearing. Analogously to the $H$- and
$L$-resonances, the radius where
\be r_{\mu \tau}=\left(\frac{\varepsilon 2E_\nu\sqrt{2}G_FY_e\lambda_r}{\Delta
m^2_{atm}} \right)^{1/3}R \label{e:resomutau}
\ee
 defines approximatively the $\mu\tau$-resonance~\cite{Akhmedov:2002zj}. We also define here
\be
 \Delta r_{\mu\tau} = \tan 2\theta_{23}\left|\frac{V_{\mu
\tau}'}{V_{\mu \tau}}\right|^{-1}_{r=r_{\mu \tau}} \propto \tan
2\theta_{23}r_{\mu \tau} \label{e:reswidth}
\ee
 as the resonance width, i.e the position around $r_{\mu\tau}$ where the $\mu-\tau$ resonance effects take place.

\begin{figure}[h]
\vspace{.6cm}
\centerline{\includegraphics[scale=0.4,angle=0]{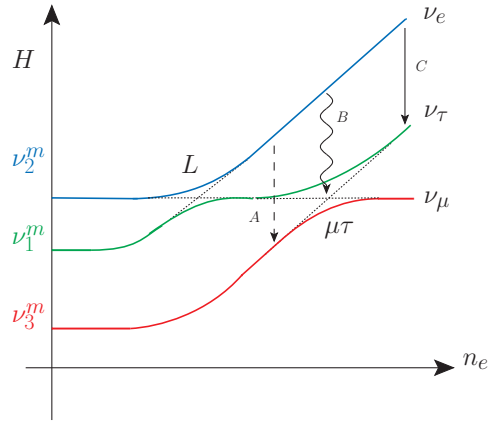}\hspace{.2cm}}
\caption{Level crossing scheme of neutrino conversion for the
inverted hierarchy in a medium with a large $V_{\mu \tau}$ with
$\theta_{23}$-mixing in the first octant.}\label{fig:theta40}
\end{figure}

In \cite{EstebanPretel:2007ec}, it has been shown that in the
single angle approximation the neutrino-neutrino interaction term
can be written as
\be
\mu(r)=\mu_0\frac{R^4}{r^4}\frac{1}{2-R^2/r^2} \ee where \be
\mu_0=\sqrt{2}G_F(F^R_{\bar{\nu}_e}-F^R_{\bar{\nu}_x})
\ee
 The bipolar transition starts when the synchronization ends
($r=r_{syn}$), and finish itself at $r_{bip}$ defined in
\cite{Hannestad:2006nj} by the relation: \be \mu(r_{bip})\simeq
\frac{|\Delta m^2_{atm}|}{2 E_\nu}. \ee On the left panel of
Fig.(\ref{fig:probaeps10-3}), for $E_\nu = 15$ MeV, the $\mu-\tau$
resonance occurs before the bipolar transition
($r_{\mu\tau}+\frac{\Delta r_{\mu\tau}}{2} < r_{bip}$).
Consequently, the matter eigenstates $\nu_1^m$ and $\nu_3^m$ start
initially as a mix of $\nu_{\mu}$ and $\nu_{\tau}$. As can be seen
in Fig.(\ref{fig:theta40}), in this case the collective effects
(represented by an arrow with a dashed line) will lead to $\nu_e
\approx \nu_3^m$ where $\nu_3^m$ is, at this moment, a mix of
$\nu_{\mu}$ and $\nu_{\tau}$. We call this type of behaviour for
the parameters $(\lambda_r,\varepsilon)$ type A, it is
equivalent to what happens in region III
in~\cite{EstebanPretel:2009is}. In this region, as in the Standard
Model case, the exact place where the L-resonance occurs does not
really modify the value of $<P(\nu_e\rightarrow \nu_e)>_{\rm{exit}}$.

On the right panel of Fig.(\ref{fig:probaeps10-3}), the $\mu-\tau$
resonance occurs right after the bipolar transition ($r_{\mu\tau}-\frac{\Delta r_{\mu\tau}}{2}
\gtrsim r_{bip}$). As can be seen in Fig.(\ref{fig:theta40}), in
this case the collective effects (represented by an arrow with a
full line) will imply that $\nu_2^m \approx \nu_\tau$ after the transition. One would
have a perfect bipolar transition from $\nu_2^m$ to $\nu_1^m$, if
the $\mu-\tau$ resonance were well separated from the
collective effects. The type C behaviour is equivalent to what happens for parameters $(\lambda_r, \varepsilon)$ in region IVa
in~\cite{EstebanPretel:2009is}. When the $\mu-\tau$ resonance
happens, it induces an exchange between the P$(\nu_e\rightarrow
\nu_{\mu})$ and P$(\nu_e\rightarrow \nu_{\tau})$ probabilities,
which corresponds to an adiabatic evolution of $\nu_1^m$ and
$\nu_3^m$, respectively ($\approx \nu_e$ and $\approx \nu_\mu$
before the resonance). For this reason, at the L-resonance,
P$(\nu_e\rightarrow \nu_e)$ strongly increases while
P$(\nu_e\rightarrow \nu_{\mu})$ and P$(\nu_e\rightarrow
\nu_{\tau})$ decrease.

On the middle panel of Fig.(\ref{fig:probaeps10-3}), the
$\mu-\tau$ resonance occurs during the bipolar transition. We call
this behaviour type B and it is a transition behaviour between A and C.
Concerning the position of the bipolar transition w.r.t. the
$\mu-\tau$ resonance, we have the following inequalities:
$r_{\mu\tau}-\frac{\Delta r_{\mu\tau}}{2} < r_{bip} <
r_{\mu\tau}+\frac{\Delta r_{\mu\tau}}{2}$. Depending on the exact
position where the $\mu-\tau$ resonance happens, the bipolar
transition will exchange $\nu_2^m\approx \nu_e$ mostly either with
$\nu_1^m\approx \nu_{\tau}$ if $r_{bip}$ is closer to
$r_{\mu\tau}-\frac{\Delta r_{\mu\tau}}{2}$ or with $\nu_3^m\approx
\nu_{\mu}$ if $r_{bip}$ is closer to $r_{\mu\tau}+\frac{\Delta
r_{\mu\tau}}{2}$. Such "interference" between the bipolar
transition and the $\mu-\tau$ resonance is represented on
Fig.(\ref{fig:theta40}) by a wiggled arrow going from $\nu_2^m$ to
a mix of $\nu_1^m$ and $\nu_3^m$.

Note that the type of behaviour that neutrinos undergo depends on the parameter couple value
 $(\lambda_r, \varepsilon)$ but also on the value of the initial neutrino luminosity that determines the strength of the collective effects.
Consequently, for different values of parameters $\lambda_r, \varepsilon$, etc... the energies and/or the place at whom those behaviours occur will be modified.

\section{Effects of the value of $\varepsilon$}
\label{sec:effectepsilon}

In this section we study the influence of the value of
$\varepsilon$, all other parameters being fixed to their respective values
used in the previous section. In a previous paper
\cite{Gava:2009gt} we showed that, according to certain value of
SUSY parameters, the radiative correction terms could yield values
of $\varepsilon$ typically contained between $\simeq
5.4\times10^{-5}$ i.e the SM value and $\simeq 2\times10^{-2}$.
Obviously the parameters space has not been fully scanned and
larger value of $\varepsilon$ might possibly be found. Modifying
this value will of course shift the position (see
Eq.\ref{e:resomutau}) where the $\mu-\tau$ resonance occurs.
Consequently, the energy values for a $\nu_e$ to have a behaviour
type A, B or C will be different.

Because of the particular form of the $\nu_e$ flux in the inverted
hierarchy due to the neutrino-neutrino interaction, i.e the
spectral split phenomenon \cite{Raffelt:2007cb,Raffelt:2007xt},
the different value of $\varepsilon$ may influence the flux in
very different ways. In the previous section, we have identified
three different types of behaviour for a $\nu_e$ going through the
supernova. Since the $\mu-\tau$ resonance obviously depends on the
value of the density and on the value of $\varepsilon$, we can
define a general quantity $N_{\mu \tau}$ equal to $\lambda_r
\varepsilon = \frac{V_{\mu \tau}}{\sqrt{2}G_F} $. In this section,
we separate the energy spectra of $\nu_e$ (received on Earth) in
three typical regions of $N_{\mu \tau}$ associated to typical
behaviours. Here, $\varepsilon$ is supposed to be positive and the
luminosity is $L_{\nu,r}$.

\subsection{Case I}

\begin{figure}[h]
\vspace{.6cm}
\centerline{\includegraphics[scale=0.3,angle=0]{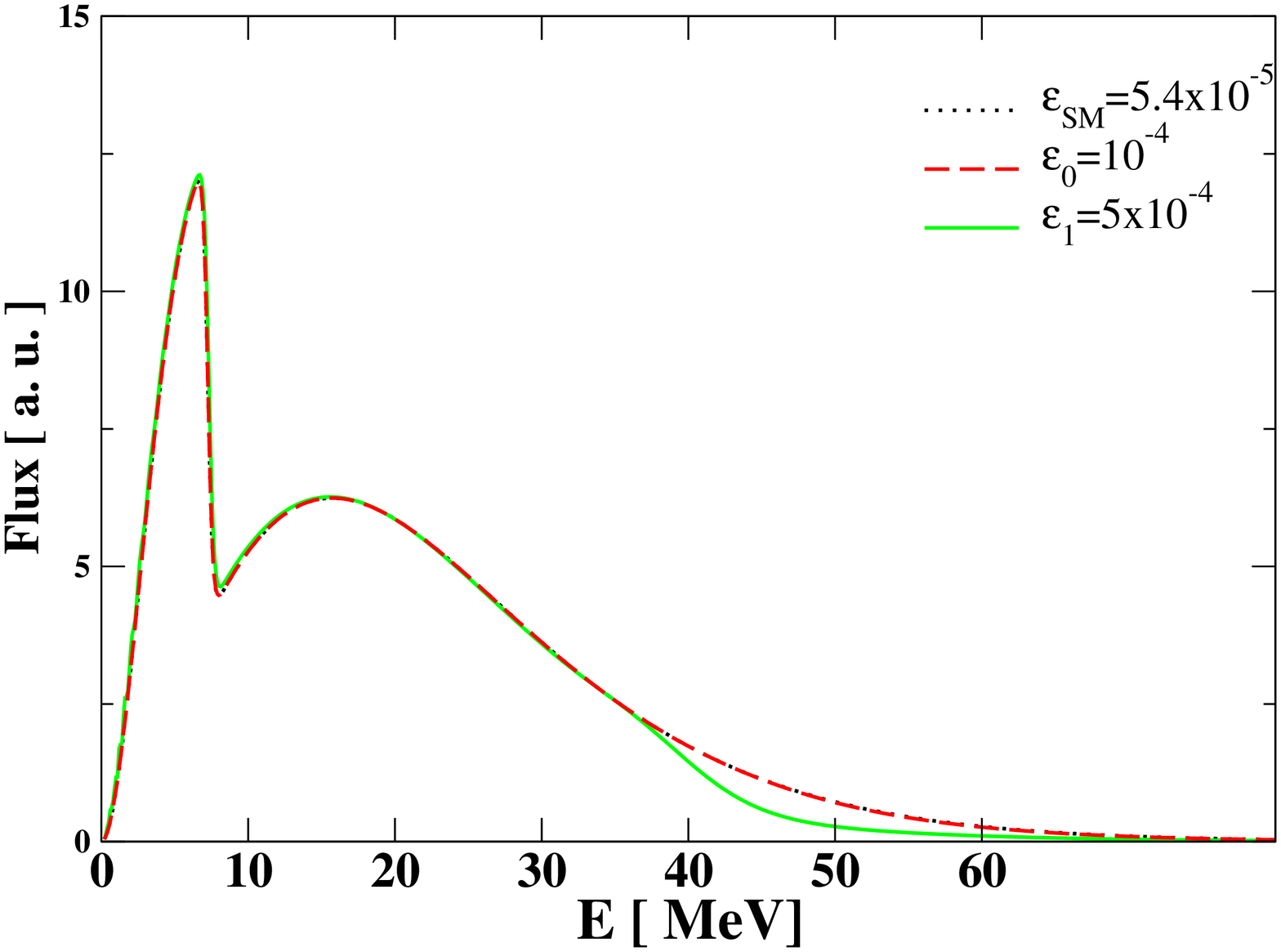}\hspace{.2cm}
\includegraphics[scale=0.3,angle=0]{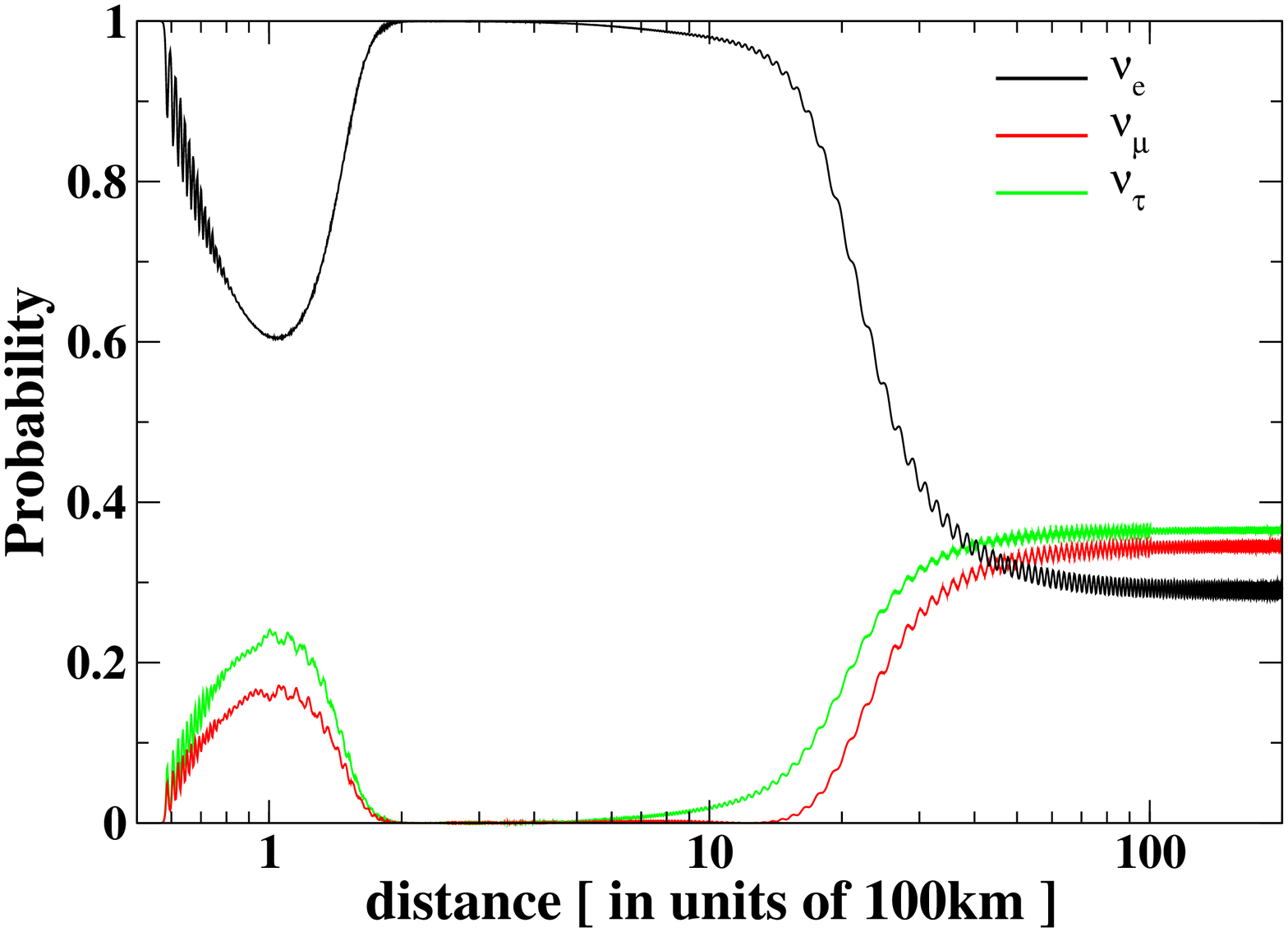}}
\caption{Left figure: $\nu_e$ flux on Earth as a function of the
energy for three different values of $\varepsilon$.
Right figure: Oscillation probabilities as a function of the
distance from the neutrinosphere with $\varepsilon = 5.4 \times
10^{-5}$ and $E_\nu=5.7$ MeV.} \label{fig:zone1}
\end{figure}

\begin{figure}[H]
\vspace{.6cm}
\centerline{\includegraphics[scale=0.35,angle=0]{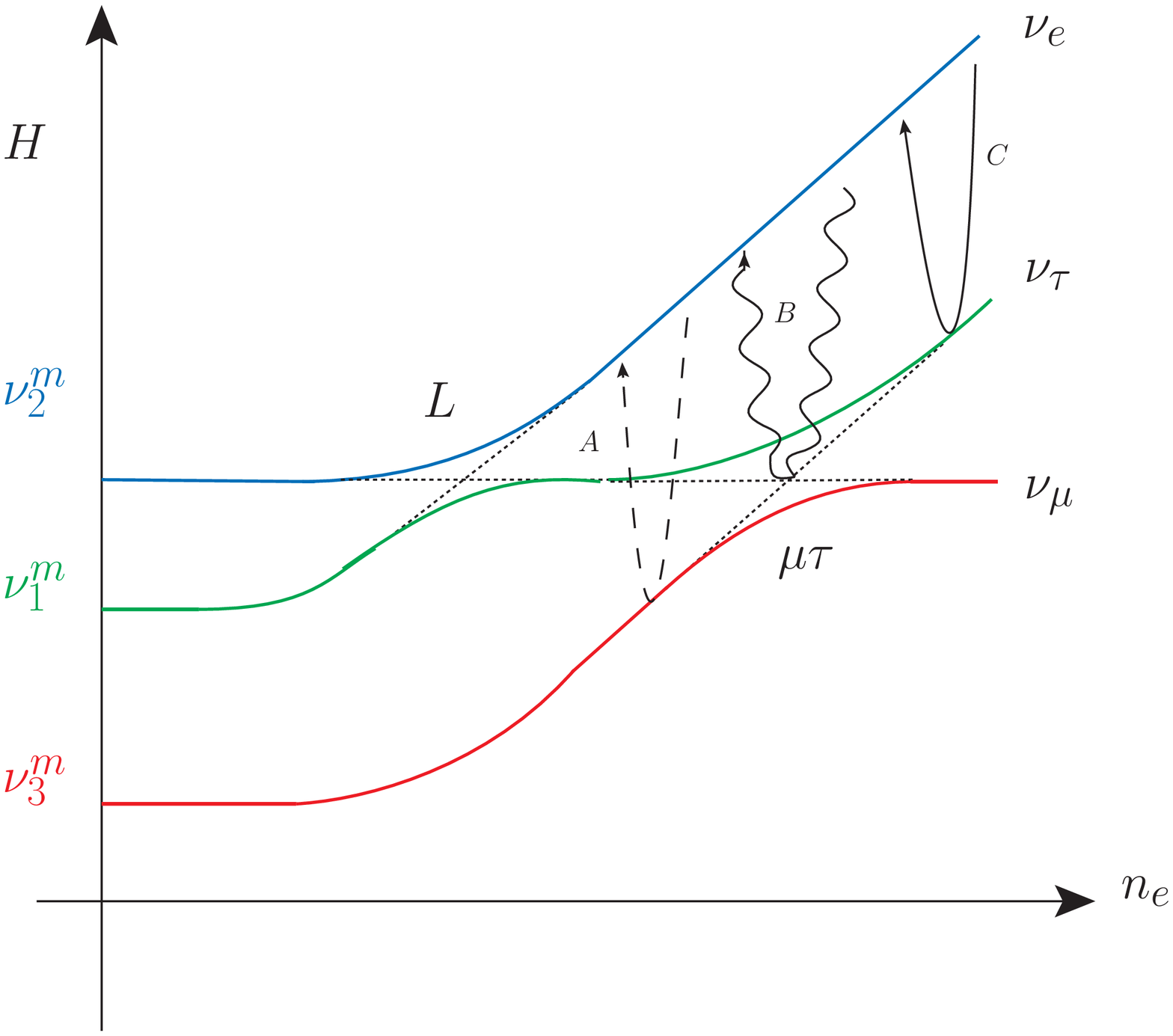}\hspace{.2cm}}
\caption{Level crossing scheme of neutrino conversion for the
inverted hierarchy in a medium with a large $V_{\mu \tau}$ with
$\theta_{23}$-mixing in the first octant and an neutrino energy
$E_\nu< E_c$.}\label{fig:theta40sp}
\end{figure}

We can define this case for $\nu_e$s that undergo a type C behaviour only for an energy above $E_\nu \simeq 50$ MeV.
This is equivalent to $N_{\mu \tau} \lesssim 1.5 \times 10^{6}$ g.cm$^{-3}$.
 For our density, the upper bound corresponds to $\varepsilon \simeq 5 \times 10^{-4}$. We show in
Fig.(\ref{fig:zone1}) three examples for $\varepsilon = 5.4 \times 10^{-5}$, $10^{-4}$ and $5 \times 10^{-4}$.
\\In the SM case, for the displayed energies, no transition can be
seen for the flux in Fig.\ref{fig:zone1}.\\
 For $\varepsilon = 10^{-4}$, the transition cannot be seen since it is above $E_\nu
\simeq 50 $ MeV. Indeed, at this energy, though $<P(\nu_e\rightarrow \nu_e)>_{\rm{exit}}$ is modified by the $\mu-\tau$ resonance, the
difference between the initial $\nu_e$ and
$\nu_x$ fluxes is too small to produce sizeable effects on the
final $\nu_e$ flux on Earth (See Eq.(\ref{e:nueflux})).\\ For
$\varepsilon = 5 \times 10^{-4}$, a type B behaviour appears at
$E_\nu \gtrsim 36 $MeV whereas the type C behaviour is reached for
$E_\nu\simeq 50$ MeV.\\ In the right part of Fig.(\ref{fig:zone1}),
we show the oscillating probabilities for an energy below $E_c$ in
the SM case. $E_c$ is known as the spectral split energy and here $E_c \simeq 8$ MeV. Looking at the electron neutrino
survival probability, we see clearly the salient features of the
$\nu-\nu$ interaction: the synchronization regime (for $r \lesssim
60$km), a bipolar transition until the spectral split climb-up
occurs (for $r \simeq 100$km), which makes the probability goes
back up to $\simeq 1$ (for $r \simeq 200$km). For this value of
$N_{\mu \tau}$ and the energy considered, the $\mu \tau$ resonance
position is well below the synchronization regime. Therefore, only
type A behaviour occurs. The presence of $V_{\mu \tau}$ can only
be seen through the fact that the bipolar transition occurs for
both $\nu_\mu$ and $\nu_\tau$ neutrinos as consistent with
Fig.(\ref{fig:theta40sp}) and the dashed arrow A. In conclusion, in this region of
$N_{\mu \tau}$, the $\mu-\tau$ resonance happens before the bipolar region, hence the $\nu_e$s
undergo a type A behaviour as defined above for $E_\nu \lesssim 36 $MeV.

\subsection{Case II}

\begin{figure}[H]
\vspace{.6cm}
\centerline{\includegraphics[scale=0.3,angle=0]{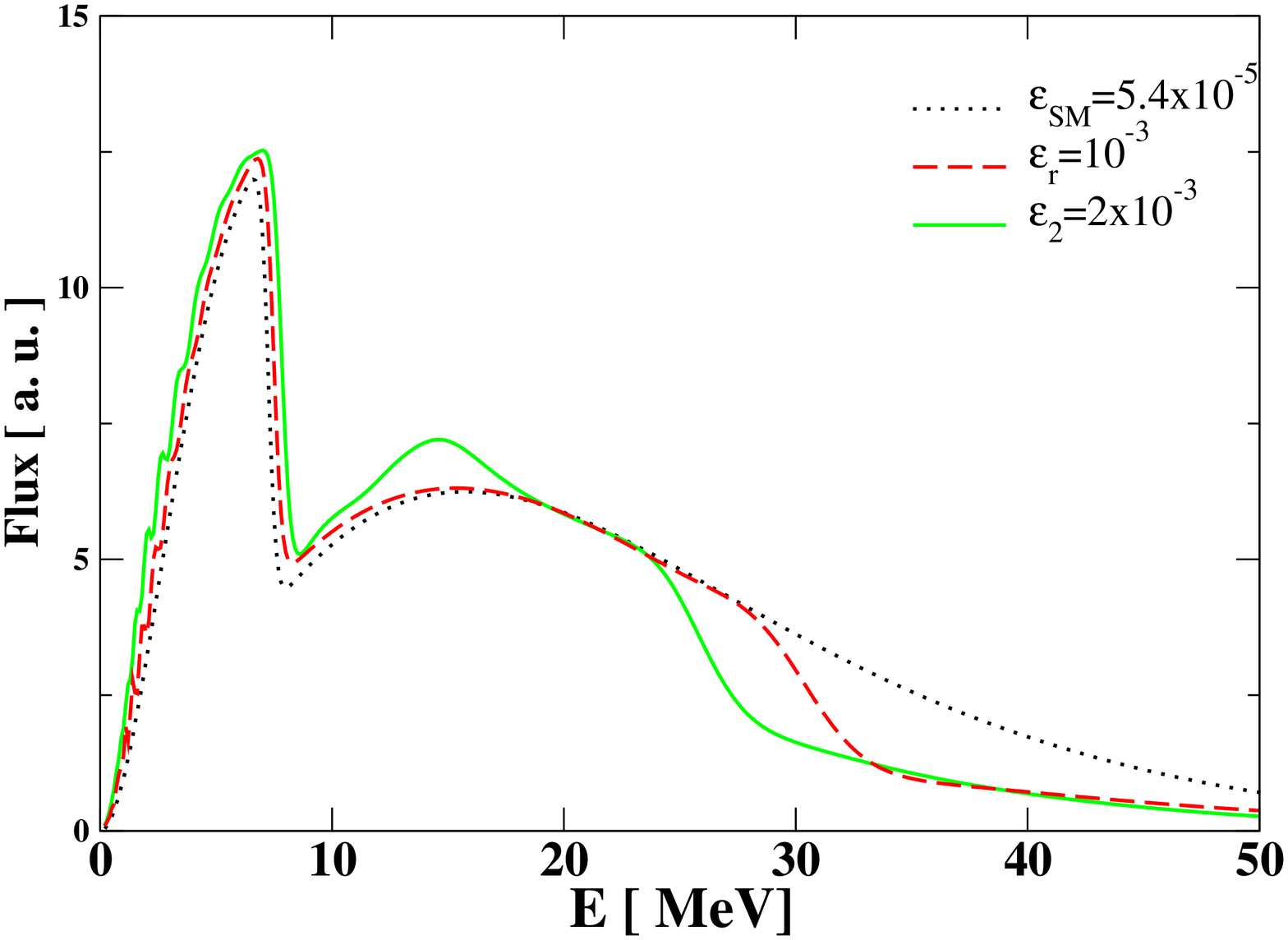}\hspace{.2cm}
\includegraphics[scale=0.3,angle=0]{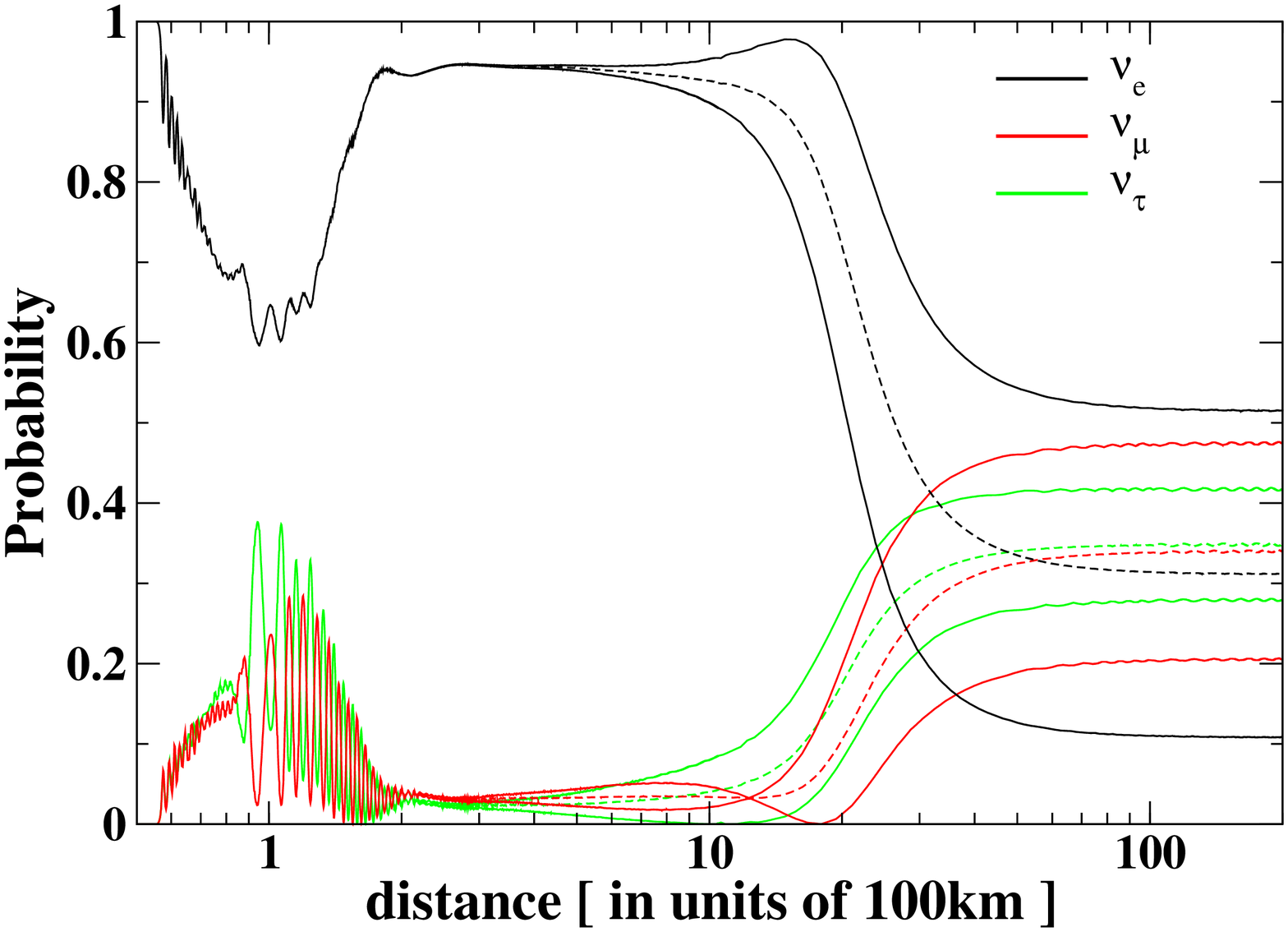}}
\caption{Left figure: $\nu_e$ flux on Earth as a function of the
energy for three different values of $\varepsilon$.
Right figure: Oscillation probabilities as a function of the
distance from the neutrinosphere with $\varepsilon = 2 \times
10^{-3}$ and $E_\nu=5.7$ MeV. We show the average as dashed lines and the envelopes of the fast-oscillating curves as solid lines.} \label{fig:zone2}
\end{figure}

We can define this case for $\nu_e$s that undergo a type B
transition for energies between $E_\nu \simeq E_c$ and $E_\nu \simeq
50$ MeV. Such behaviour corresponds to the interval $1.5 \times
10^{6}$ g.cm$^{-3} \lesssim N_{\mu \tau} \lesssim 2 \times
10^{7}$g.cm$^{-3}$. For our density, the upper bound corresponds
to $\varepsilon \simeq 7\times10^{-3}$ while the lower bound is
associated to $\varepsilon\simeq 5 \times 10^{-4}$. We show in
Fig.(\ref{fig:zone2}) two examples of such large $\varepsilon$
value: $\varepsilon = 10^{-3}$ and $2 \times 10^{-3}$. We also
show, for comparison purpose, the $\nu_e$ flux for the SM value of
$\varepsilon$.
For $\varepsilon = 10^{-3}$, the transition zone appears for $28$
MeV $\lesssim E_\nu \lesssim 33$ MeV.\\ For the case $\varepsilon = 2 \times 10^{-3}$,
$P(\nu_e \rightarrow \nu_e)$ leaves a type A behaviour for $E_\nu
\gtrsim 10$ MeV and the type C behaviour is reached for $E_\nu
\gtrsim 30$ MeV. In this case, we therefore have a larger
transition zone, which is due to the fact that the resonance width
is bigger. Indeed, this resonance width being proportional to
$r_{\mu \tau}$, it varies like $(\varepsilon E
\lambda_r)^{\frac{1}{3}}$, and therefore gets larger when
$\varepsilon$ goes from 10$^{-3}$ to 2 $\times 10^{-3}$. This is
why the flux moves away from the $\nu_x$ initial flux\footnote{The
initial $\nu_e$ and $\nu_x$ flux can be seen in
Fig.(\ref{fig:zone3}).}
 for the $E_\nu \sim 15$ MeV, i.e
$<P(\nu_e\rightarrow \nu_e)>_{\rm{exit}}$ is close but
not equal to zero. For instance, at $E_\nu=15$ MeV, $<
P(\nu_e\rightarrow \nu_e)>_{\rm{exit}}$=0.15 which implies, recalling
Eq.(\ref{e:nueflux}), $\Phi_{\nu_{e}}=
0.15\times(F^R_{\nu_{e}}-F^R_{\nu_{\mu}})+F^R_{\nu_{\mu}}$. For $E_\nu
\sim 20$ MeV, it seems, in a misleading way that the $<$
P$(\nu_e\rightarrow \nu_e)$ $>_{\rm{exit}}$ has gone down to a zero value
again. Actually, this is due to the fact that, for this energy,
the initial fluxes of $\nu_e$ and of $\nu_x$ cross. Consequently,
the $\nu_e$ flux on Earth is equal to the initial flux of $\nu_x$
for $E_\nu \sim 20$ MeV. Indeed, we can verify that the type B behaviour starts anew for $20$ MeV $< E_\nu < 30$ MeV.

In the right part of Fig.(\ref{fig:zone2}), we show the
oscillating probabilities for an energy below $E_c$. For this
value of $N_{\mu \tau}$, the $\mu \tau$ resonance position is
mainly below the bipolar transition. Nevertheless, it leads to a
small disruption of the spectral split climb-up since the electron
neutrino survival probability does not goes up back to a value of $\simeq 1$
but goes to $\simeq 0.93$ instead. $<P(\nu_e
\rightarrow \nu_e)>_{\rm{exit}}$ is slightly above the one yielded in case I.
This explains the non perfect superimposition of the flux for
energy below $E_c$. Note that the precise interplay between the
$\mu \tau$ resonance and the neutrino-neutrino interaction in this
case should be investigated more deeply.

\subsection{Case III}

\begin{figure}[H]
\vspace{.6cm}
\centerline{\includegraphics[scale=0.3,angle=0]{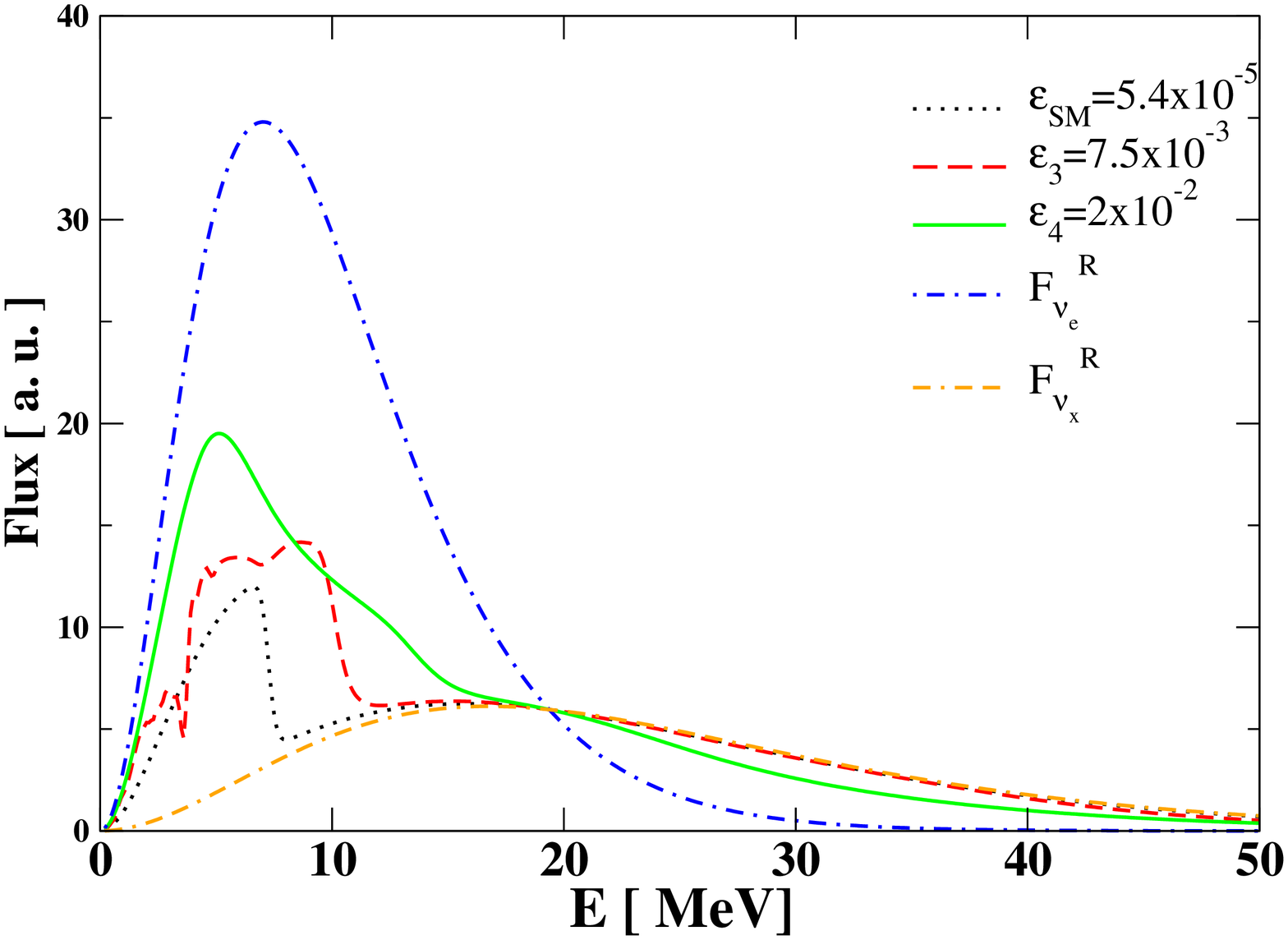}\hspace{.2cm}
\includegraphics[scale=0.3,angle=0]{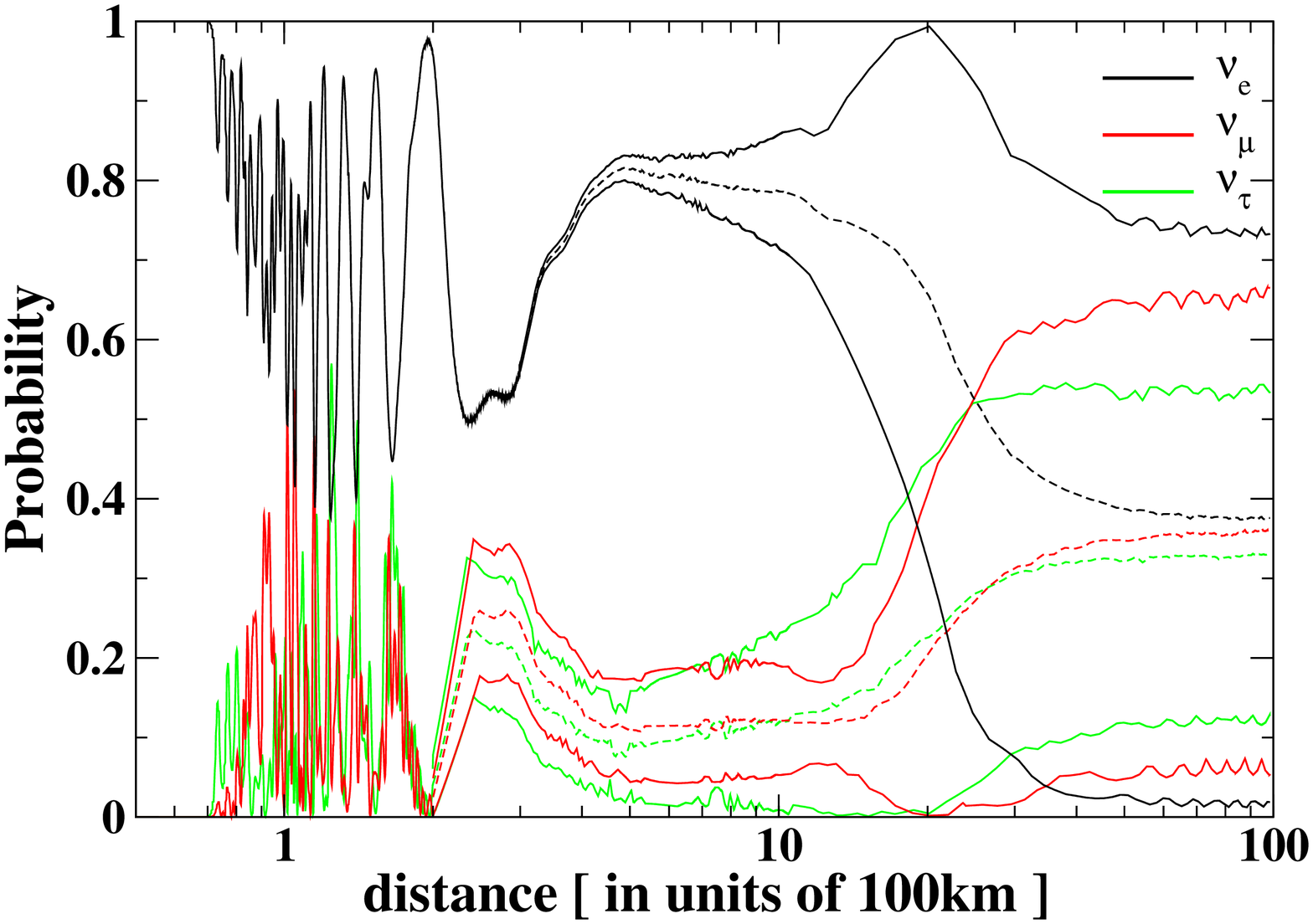}}
\caption{Left figure: $\nu_e$ flux on Earth as a function of the
energy for three different values of $\varepsilon$. We also show here the initial $\nu_e$ and $\nu_x$ flux emitted at the neutrinosphere.
Right figure: Oscillation probabilities as a function of the
distance from the neutrinosphere with $\varepsilon = 7.5 \times 10^{-3}$ and $E_\nu=5.7$ MeV. We show the average as dashed lines and the envelopes of the fast-oscillating curves as solid lines.} \label{fig:zone3}
\end{figure}

We can define this case by the interval $2 \times
10^{7}$g.cm$^{-3} \lesssim N_{\mu \tau} \lesssim 6 \times
10^{7}$g.cm$^{-3}$. For our density, the upper bound corresponds
to $\varepsilon \simeq 2 \times10^{-2}$ while the lower bound is
$\varepsilon \simeq 7 \times10^{-3}$. We show in
Fig.(\ref{fig:zone3}) two examples for $\varepsilon = 7.5 \times
10^{-3}$ and $2 \times 10^{-2}$ exhibiting two different cases. Note
that in the case III, $\Delta r_{\mu\tau}$ is significantly
greater than in the previous cases I and II because it varies as
$\propto r_{\mu\tau}$.

In the case III, the $\mu-\tau$ resonance goes out of the
synchronized region for $E \lesssim E_c$ where $E_c$ is the energy
where the split occurs. Thus it will create a disruption of the
spectral split phenomenon.\\ In the case $\varepsilon = 7.5 \times
10^{-3}$, we surprisingly see that a type A behaviour occurs for $E_c \lesssim E_\nu \lesssim$ 40 MeV. It turns out that the bipolar transition for theses energies is extremely disrupted. This may be due to a non-linear interference between the bipolar transition and the width of the $\mu -\tau$ resonance but this should deserve further investigation. For $E \gtrsim$ 40 MeV, we begin a transition phase.
\\In the case $\varepsilon = 2 \times 10^{-2}$, for $E_\nu
\gtrsim E_c$, $r_{\mu\tau}+\frac{\Delta r_{\mu\tau}}{2} >
r_{bip}$, this corresponds to the beginning of the transition
phase which lasts till $r_{\mu\tau}-\frac{\Delta r_{\mu\tau}}{2} >
r_{bip}$. It is important to note that the more the energy
increases the larger the resonance width is, so the phase transition begins for $E_\nu
\lesssim E_c$. Note that the spectral split signature is vanishing
for this value of $\varepsilon$.

The disruption of the spectral split phenomenon is explicitly shown in the right part of Fig.(\ref{fig:zone3}).
 The bipolar transition shows huge oscillations due to a powerful interplay with the $\mu \tau$ resonance,
  while the spectral split climb-up makes the electron survival probability go up only to a value of $\simeq 0.8$ (see right part of Fig(\ref{fig:zone3})) contrary to the SM case where it goes back up to $\simeq 1$.
    Eventually, this leads to a higher value of $<P(\nu_e\rightarrow \nu_e)>_{\rm{exit}}$
     which was seen on the left part of Fig.(\ref{fig:zone3}). This phenomenon is represented in Fig.(\ref{fig:theta40sp}) by the wiggled arrow B.

Finally, note that in this section, since we studied the influence of an increasing value of $\varepsilon$, we have displayed in the three cases the oscillating probabilities with $E_\nu <E_c$. Indeed, as $\varepsilon$ increases, the resonance occurs for lower and lower energies since the density remains in the three cases constant. This can be seen in Eq.(\ref{e:resomutau}). It is therefore interesting to look at the evolution of the probabilities at the energy around which the $\mu-\tau$ resonance occurs. The evolution probabilities for $E_\nu>E_c$ can be seen in Figs. (\ref{fig:probaeps10-3}).
\section{The degeneracy of $\varepsilon$ with respect to other parameters}
\label{sec:parameters}

As seen in Sec. \ref{sec:typical}, the typical behaviour A, B or C
are the consequence of a certain position of the $\mu-\tau$
resonance in comparison with the bipolar transition. In the
following subsections, we study in particular one parameter that
influence either the resonance or the $\nu-\nu$ interaction, the
other being fixed to their reference value as defined in Sec.
\ref{sec:theoframe}.

\subsection{Effects of $\theta_{23}$ and of the sign of $\varepsilon$}
\label{sec:effecttheta}

Changing the value of $\theta_{23}$ in such way that it changes of
octant, will automatically prevent a $\mu-\tau$ resonance as
discussed in \cite{EstebanPretel:2007yq,EstebanPretel:2009is}.
Using the two flavour MSW condition, \be \frac{\Delta
m^2_{atm}}{2E_\nu}\cos2\theta_{23}=V_{\mu \tau}=\varepsilon
\sqrt{2}G_FY_e\rho_B, \label{twoflavmutaua} \ee one can see that
changing the sign of $\varepsilon$ could be immediately
interpreted as perfectly equivalent to changing of octant.
However, in the case of a negative $\varepsilon=-|\varepsilon|$, looking at the matter interaction Hamiltonian, we have:  \be
H_m=diag(V_e, 0, -|V_{\mu \tau}|)= diag(V_e +|V_{\mu
\tau}|,|V_{\mu \tau}|, 0). \label{e:negeps} \ee In this case, the
term $|V_{\mu \tau}|$ becomes a matter potential for $\nu_\mu$
which can be seen like an effective presence of muon in the
supernova whereas the electron matter potential appears slightly
modified of the order of $\varepsilon$, i.e $2\%$ at most.

\begin{figure}[H]
\vspace{.6cm}
\centerline{\includegraphics[scale=0.3,angle=0]{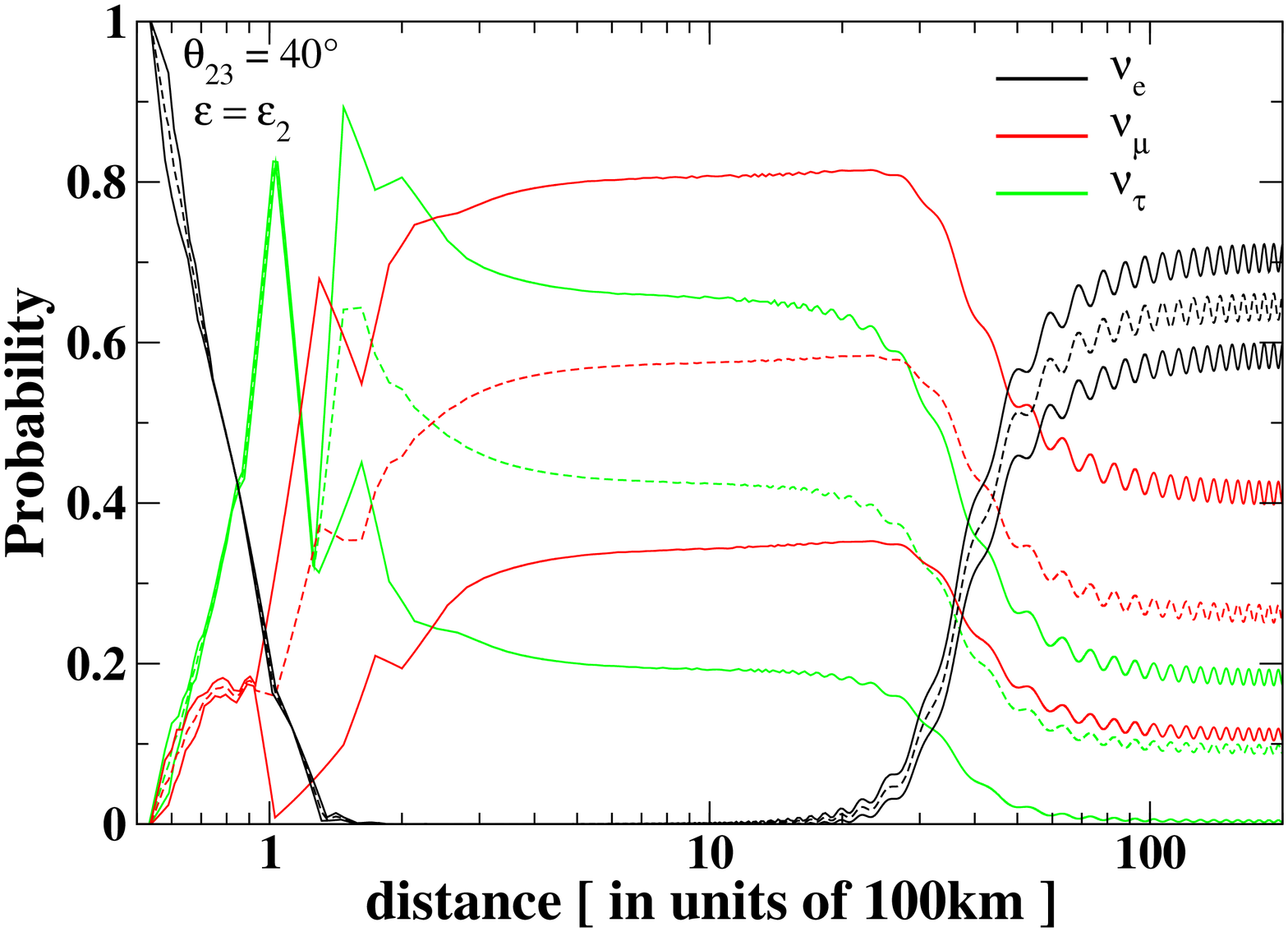}\hspace{0.25cm}\vspace{1.0cm}
\includegraphics[scale=0.3,angle=0]{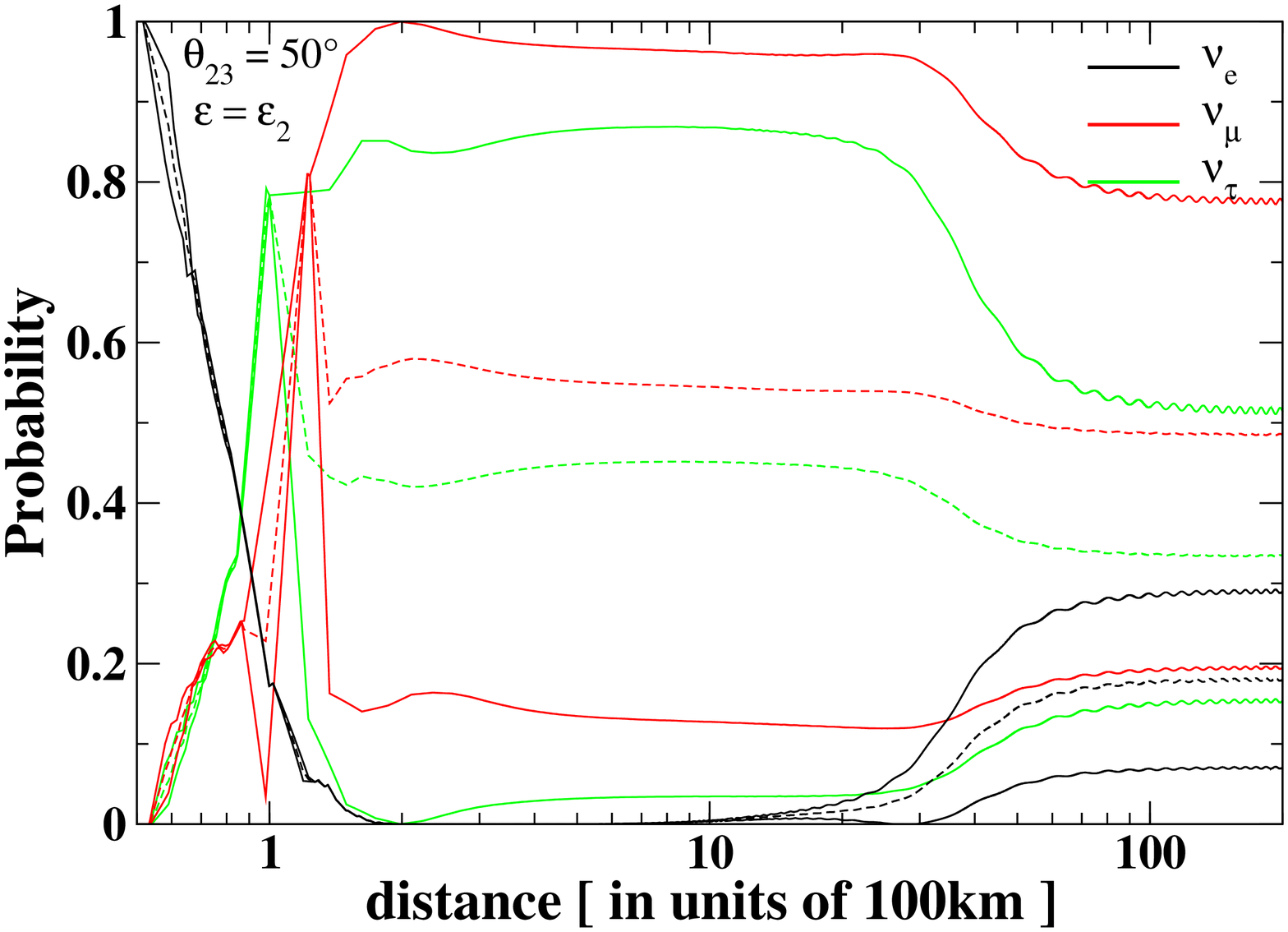}}
\centerline{
\includegraphics[scale=0.3,angle=0]{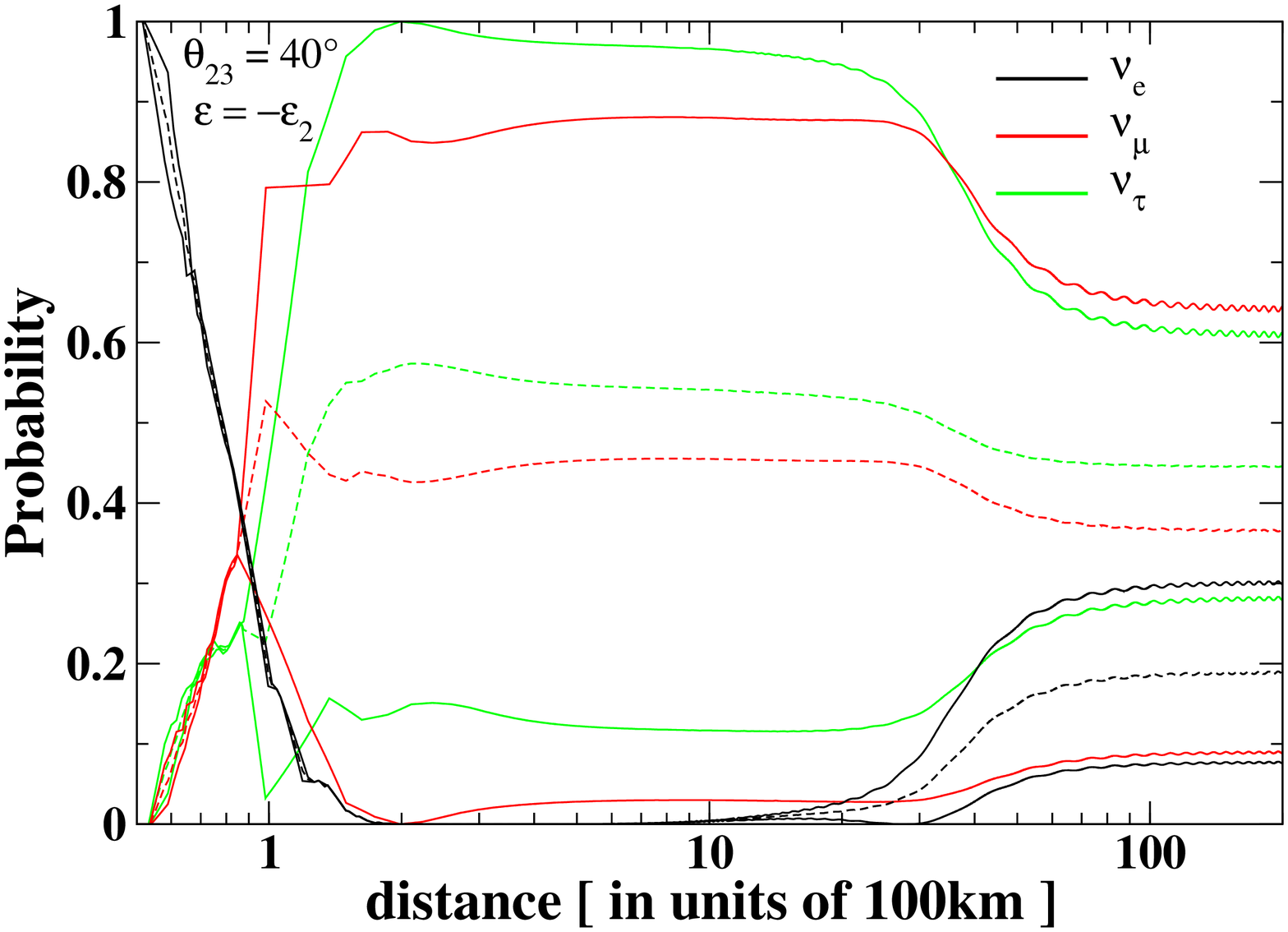}\hspace{0.25cm}\vspace{1.0cm}
\includegraphics[scale=0.3,angle=0]{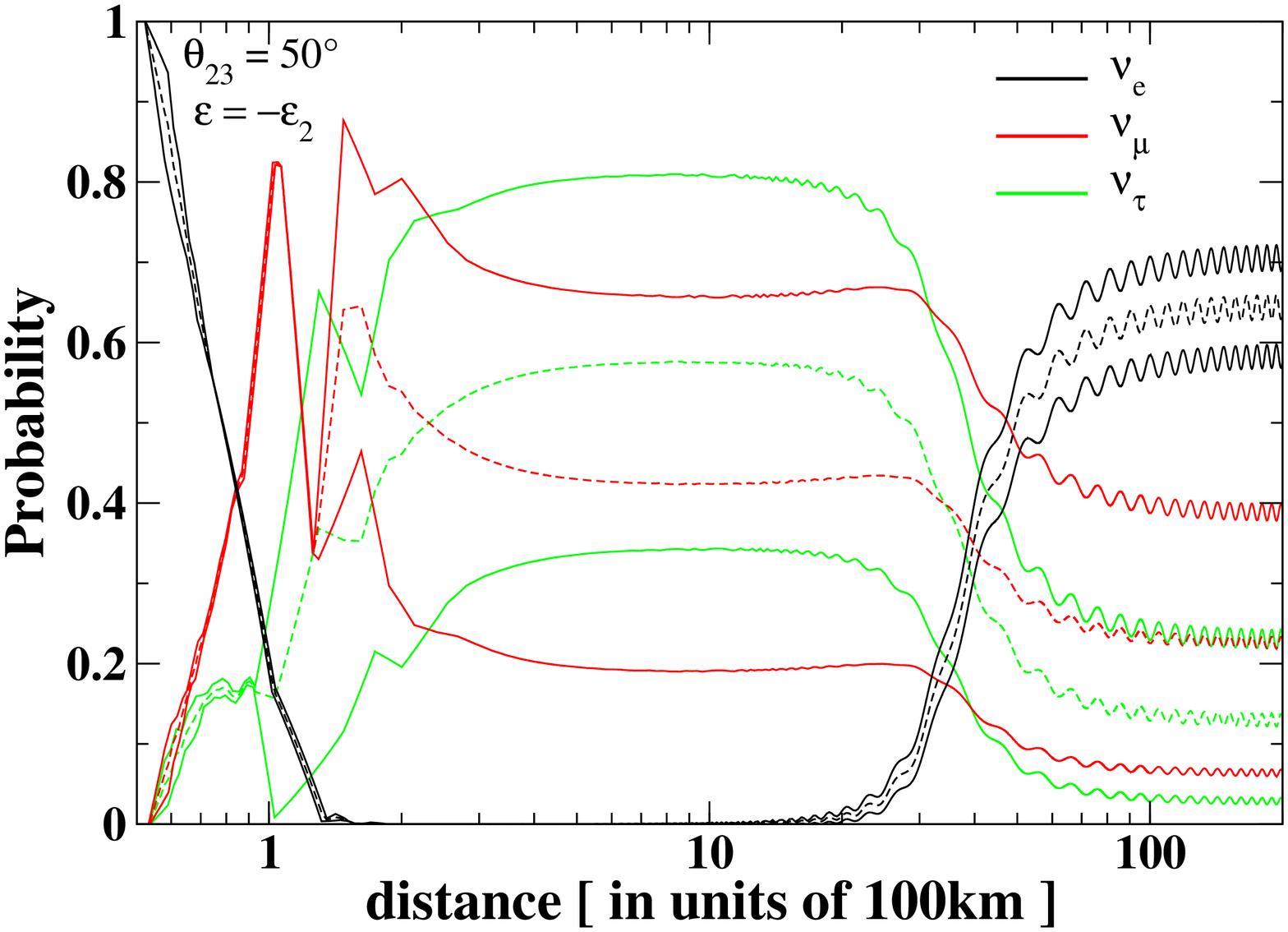}}
\caption{Oscillation probabilities as a function of the distance
from the neutrinosphere. We show the average as dashed lines and the envelopes of the fast-oscillating curves as solid lines. The figures correspond to the value of
the couple of parameters ($\varepsilon = \pm \varepsilon_{2}$,
$\theta_{23}=40^\circ$ or $50^\circ$). $\varepsilon_2$ is still equal to $2\times 10^{-3}$. In these figures, the neutrino energy is 30 MeV.}
\label{fig:probasigntheta23}
\end{figure}

This explains why, looking at the left figure panel in
Fig.(\ref{fig:probasigntheta23}), the top and lower part show
opposite behaviours for $\nu_\mu$ and $\nu_\tau$ in the first 100km; The same thing
happens for the right figure panel.
Even when the MSW condition is not fulfilled, a matter effect is
still present as can be seen in~\cite{EstebanPretel:2007yq}. This
is called the matter suppression effect and a transition, though
partial is happening.

\begin{figure}[H]
\vspace{.6cm}
\centerline{\includegraphics[scale=0.3,angle=0]{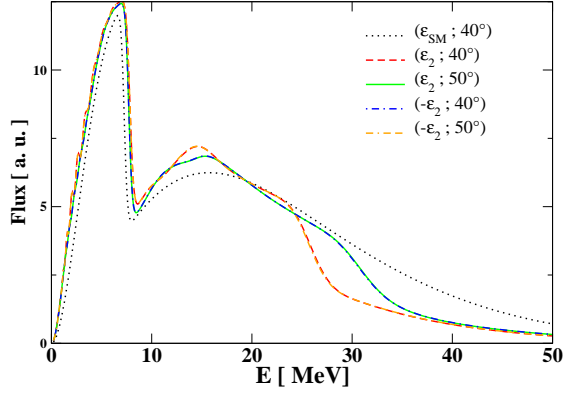}}
\caption{$\nu_e$ flux on Earth as a function of the energy for different values of the parameter couple
 ( $\varepsilon$, $\theta_{23}$). $\varepsilon_2$ is still equal to $2\times 10^{-3}$.}
 \label{fig:spspsigntheta23}
\end{figure}

\begin{figure}[H]
\vspace{.6cm}
\centerline{\includegraphics[scale=0.35,angle=0]{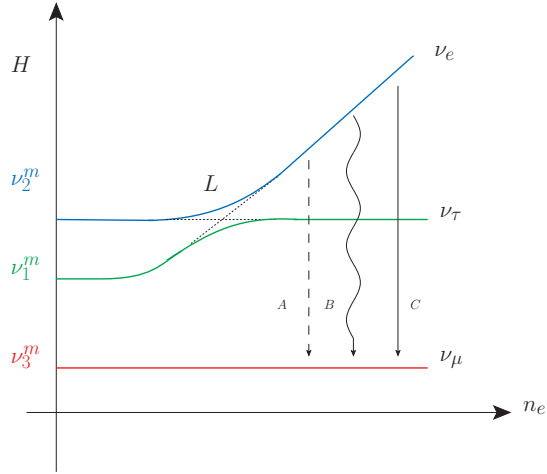}\hspace{.2cm}}
\caption{Level crossing scheme of neutrino conversion for the
inverted hierarchy in a medium with a large $V_{\mu \tau}$ with
$\theta_{23}$-mixing in the second octant.}\label{fig:theta50}
\end{figure}

In Fig.(\ref{fig:spspsigntheta23}), we observe that the energy
spectra of the electron neutrino flux on Earth, is the same for
the parameter couples ($\varepsilon>0$, $\theta_{23}=40^\circ$)
and ($\varepsilon<0$, $\theta_{23}=50^\circ$) whereas a identical
shift for the transition around 30 MeV is observed in the cases
($\varepsilon<0$, $\theta_{23}=40^\circ$) and ($\varepsilon>0$,
$\theta_{23}=50^\circ$). The fact that a similar behaviour between
the cases ($\varepsilon>0$, $\theta_{23}=40^\circ$) and
($\varepsilon<0$, $\theta_{23}=50^\circ$) is observed is due to
the fulfillment of the MSW condition (Eq.(\ref{twoflavmutaua})) in the two cases. The only
differences can be seen on the evolution with the distance of the
$\nu_\mu$ and $\nu_\tau$ oscillation probabilities, as said above
because in the first/second case, the neutrinos see
respectively an effective presence of $\tau$/$\mu$
leptons. In the opposite cases, no MSW resonance can happen but a
matter suppression effect occurs. To explain this shift in energy,
one can think the matter suppression as a transition between
$\nu_\mu$ and $\nu_\tau$ neutrinos but less effective as the MSW
resonance. Since in all cases, all the parameters except the sign
of $\varepsilon$ and the octant of $\theta_{23}$, are identical,
to maximize the effect of this matter suppression, one needs to
move it away from the collective effects region, this can be done
by increasing the neutrino energy. Note that the matter
suppression can interfere with the collective effects as
represented as a wiggled arrow in Fig.(\ref{fig:theta50}).

\subsection{Effects of the luminosity}
\label{sec:effectlum}

\begin{figure}[H]
\vspace{.6cm}
\centerline{\includegraphics[scale=0.3,angle=0]{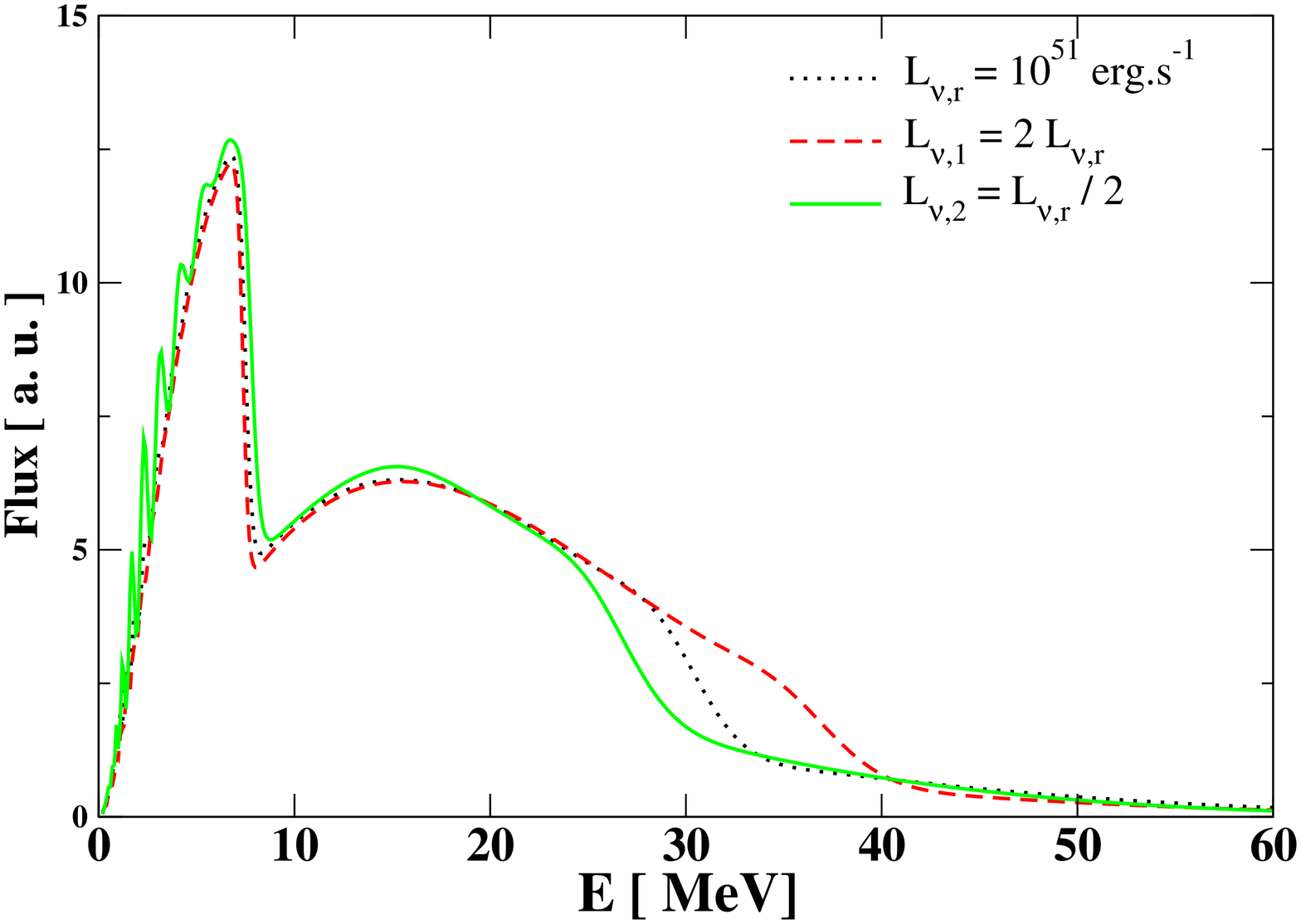}\hspace{.2cm}
\includegraphics[scale=0.3,angle=0]{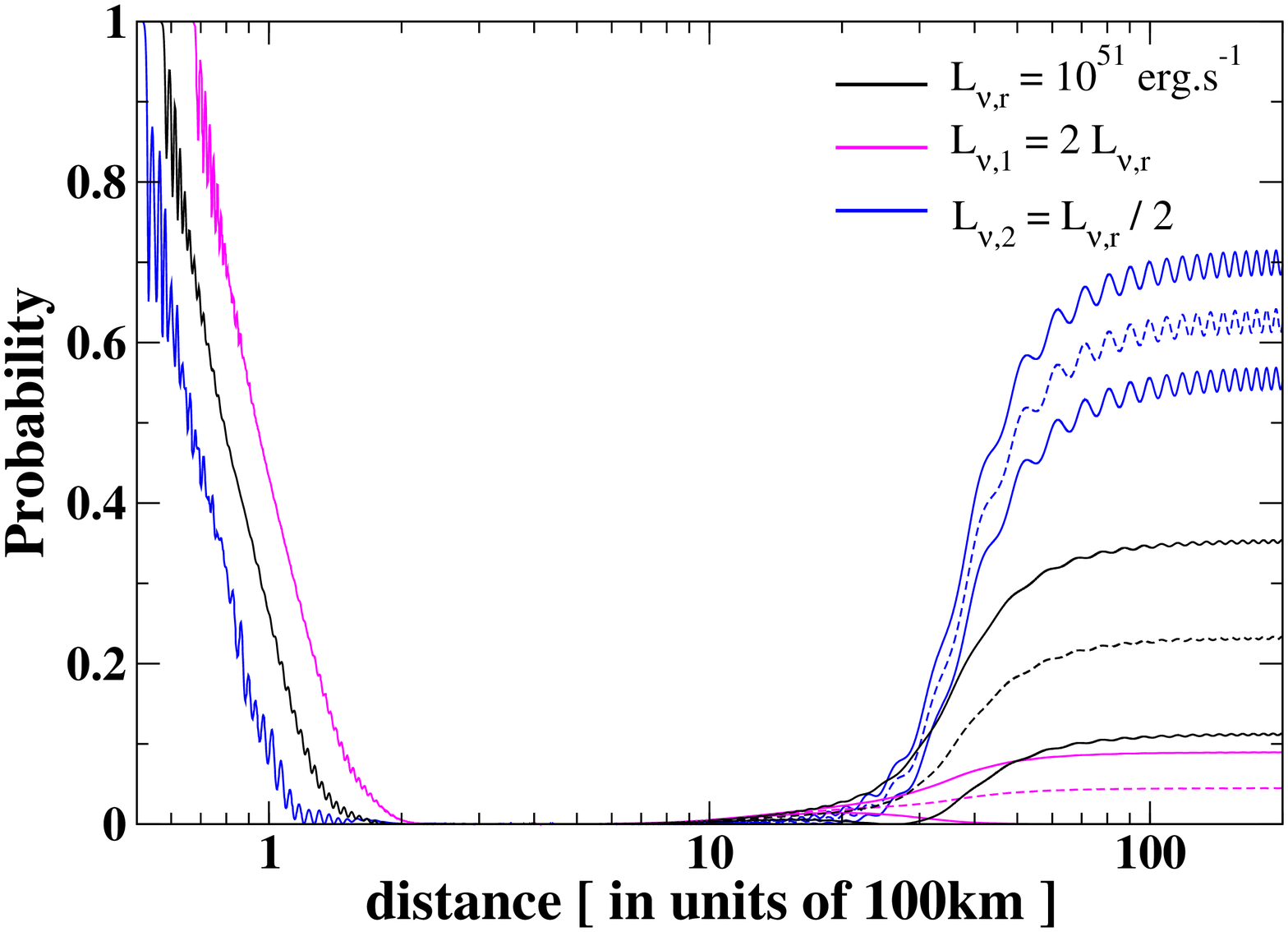}}
\caption{Left figure: luminosity-normalized $\nu_e$ flux on Earth as a function of the
energy for three different luminosities.
Right figure: Electron neutrino survival probabilities as a
function of the distance from the neutrinosphere for three different luminosities with $E_\nu=30$ MeV. We show the average as dashed lines and the envelopes of the fast-oscillating curves as solid lines.}
\label{fig:spsplumi}
\end{figure}

Effects of luminosity should not be neglected. As the cooling
phase passes, the neutrino fluxes emitted at their respective
neutrinosphere, decrease in intensity. Such diminution can be
approximated by a decreasing exponential as seen in
Eq.(\ref{e:lumexp}). Consequently, the strength of the $\nu-\nu$
interaction shrinks, implying a slow diminution of the collective
effect predominating zone: \be r_{bip}\sim \mu_0^{\frac{1}{4}}
\sim \exp\left(\frac{-t}{4 \tau}\right) \ee As the bipolar radius
decreases, $r_{\mu \tau}$ being fixed, the electron neutrino
probability behaviour tends to be like the type C behaviour. This
can be seen in Fig.(\ref{fig:spsplumi}) where the fluxes have been
normalized to the reference value in order to appreciate only the
$\mu-\tau$ / $\nu-\nu$ interplay effects. Knowing that $r_{\mu
\tau}\sim (\lambda_r \varepsilon)^{\frac{1}{3}}$, multiplying the
luminosity by a factor $\alpha$ would in principle be equivalent
to multiply either $\lambda_r$ or $\varepsilon$ by a factor
$\alpha^{\frac{-3}{4}}$ since \be \frac{r_{\mu \tau}}{r_{bip}}\sim
(\lambda_r \varepsilon)^{\frac{1}{3}}\mu_0^{\frac{-1}{4}}
\label{e:ratioradius} \ee Therefore, increasing the luminosity or
shifting inwards the position of the $\mu-\tau$ resonance is
rather equivalent until the neutrino density is too weak to
maintain collective effects.

Looking at the left part of Fig.(\ref{fig:spsplumi}), one sees the
energy spectra of $\nu_e$ varying as predicted above. When
$L_\nu=2 \times 10^{51} $erg.s$^{-1}$ the $\nu_e$ present a
similar shape with the reference case, except that the transition
zone starts for $E_\nu \simeq35$ MeV. Indeed, a higher luminosity
is equivalent to a $\mu-\tau$ resonance position deeper inside the
supernova. Therefore, neutrinos need a higher energy value to
begin to see the type B behaviour. Symmetrically, a lower value of
the initial neutrino flux will make the type B behaviour appears
for lower energies, i.e $E_\nu \simeq23$ MeV in this case. Note
that because of a low value of the neutrino flux, the collective
effects start to weaken which leads to some oscillations in the
flux for energies $E_\nu \lesssim E_c$. Equivalently, one can look
at the right part of Fig.(\ref{fig:spsplumi}) where we show
the electron neutrino survival probability for the three different
values of the luminosity: $L_{\nu,r}$, $L_{\nu,1}$ and
$L_{\nu,2}$. The value of $<P(\nu_e \rightarrow
\nu_e)>_{\rm{exit}}$ is anew highly dependent of the relative position of
$r_{\mu \tau}$ w.r.t $r_{bip}$.
 This figure shows explicitly the increase of $r_{bip}$ with the luminosity.

\subsection{Effects of the density}
\label{sec:effectdens}

\begin{figure}[H]
\vspace{.6cm}
\centerline{\includegraphics[scale=0.3,angle=0]{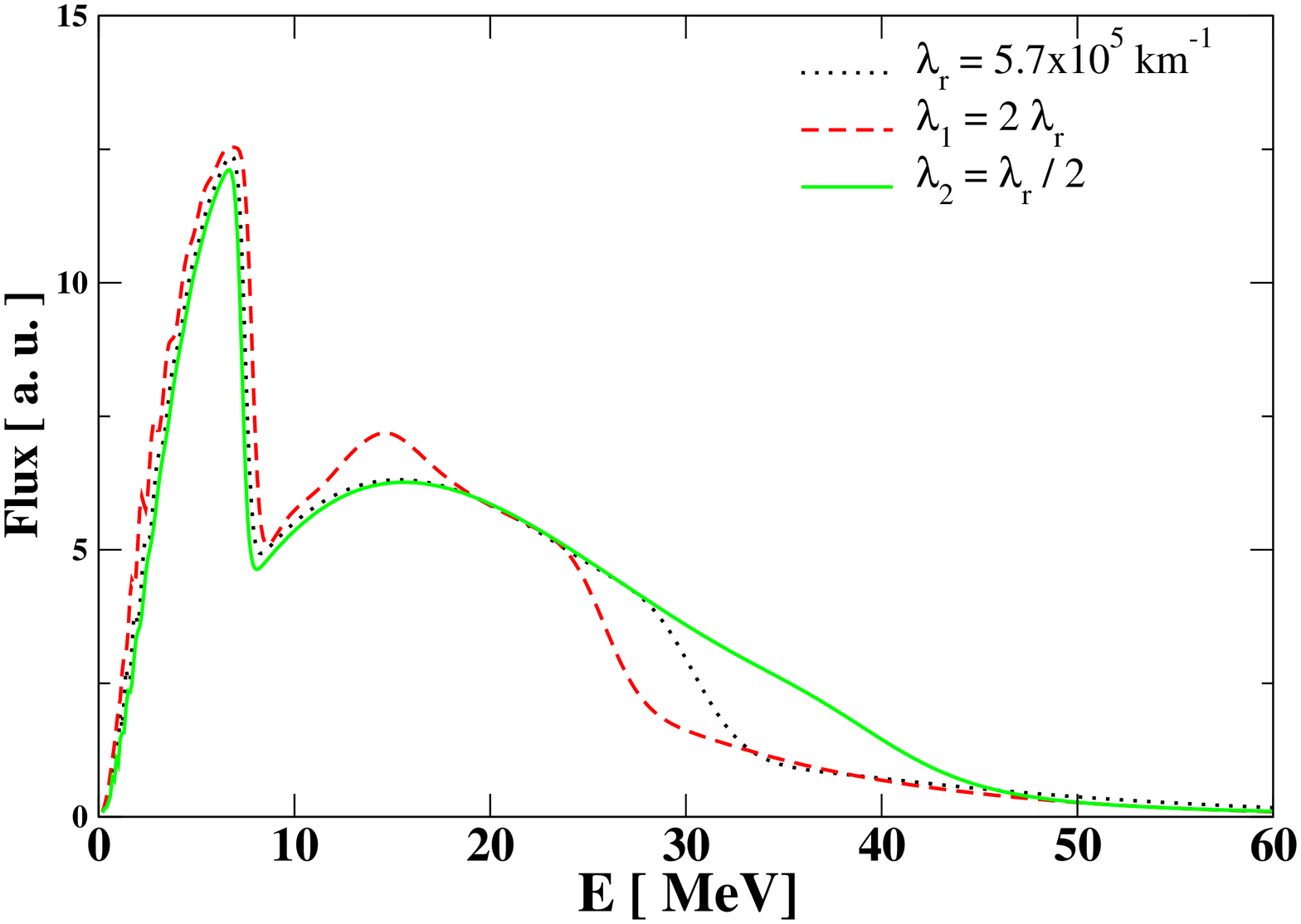}\hspace{.2cm}
\includegraphics[scale=0.3,angle=0]{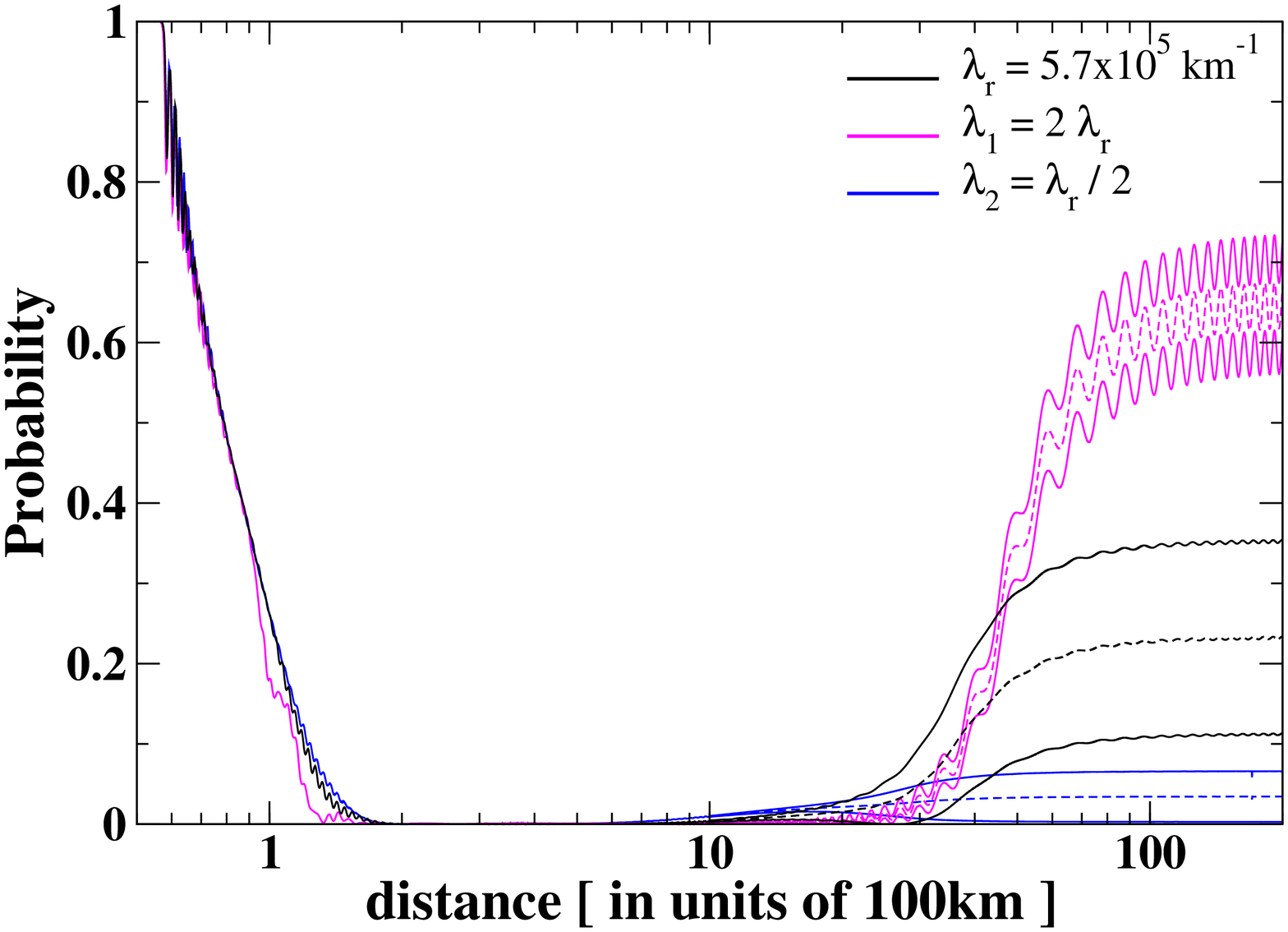}}
\caption{Left figure: $\nu_e$ flux on Earth as a function of the
energy for three different densities.
Right figure: Electron neutrino survival probabilities as a
function of the distance from the neutrinosphere for three different densities with $E_\nu=30$ MeV. We show the average as dashed lines and the envelopes of the fast-oscillating curves as solid lines.}
\label{fig:spspdensity}
\end{figure}

Depending on the mass of the progenitor star, the explosion
releases a variable energy which, under the form of a shock-wave,
can lead to important variation on the matter density, whose
precise profile remains partially unknown. In our analytical
profile, it means that the value of $\lambda_r$ will decrease with
time. The effect of the density variation through the parameter
$\lambda_r$ is crystal clear. Using Eq.(\ref{e:resomutau}), one
can see that the position of the $\mu-\tau$ resonance varies like
$\lambda_r^{\frac{1}{3}}$. Therefore, changing the density or the
value of the radiative correction term $\varepsilon$ is almost
perfectly equivalent as long as the density does not reach very
high value such as in the accretion phase. This can be seen in the
left part of Fig.(\ref{fig:spspdensity}) where
the density is $\lambda_1 = 2\lambda_r$ and $\lambda_2=
\lambda_r/2$ are respectively equivalent to the reference case but
with $\varepsilon_1=2\times 10^{-3}$ and $\varepsilon_2=5 \times
10^{-4}$. The incertitude concerning the density profile plays an
important role as $N_{\mu \tau}$ gives rise to three well distinct
behaviours corresponding to the cases I, II and III. Equivalently,
one can look at the right part of Fig.(\ref{fig:spspdensity}),
where we show the electron neutrino survival probability for the
three different values of the density: $\lambda_r$, $\lambda_1$
and $\lambda_2$; We can notice that $r_{bip}$ is the same for the
three curves, which is consistent with the fact that the
luminosity has not been modified. The bipolar transition is a bit
disrupted in the case where the matter is more dense. This is due
to the fact that the $V_{\mu \tau}$ term is larger, which implies
a more powerful interplay with the $\mu-\tau$ resonance.
 Moreover, we see a light shift of the L-resonance position towards the outside
  of the supernova with an increasing value of the density parameter $\lambda$.

\section{Detection of the signal on Earth}
\label{sec:detectEarth}

Since the explosion of SN1987A in the Large Magelanic Cloud, it
has been proved that neutrinos from supernova can be detected on
Earth. Such astrophysical event can yield a tremendous amount of
information through the neutrino fluxes received on Earth. With
the currently running neutrino observatories, a high-statistics
signal for a galactic SN explosion is expected in the electron
anti-neutrino channel. For instance, Super-Kamiokande could detect
through the reaction $\bar{\nu}_e + p \rightarrow n + e^+$ about
$\sim10^4$ events. Nevertheless, larger detectors could yield an
even better statistics and therefore could be sensitive to more
discreet features of the explosion. Also, they logically could detect a more distant explosion with
the same current number of event statistics.

Three main types of multi-purpose detectors are proposed to
enhance the quality of information about the neutrino properties
and the explosion mechanism that we will receive from the
supernova neutrino fluxes. The proposed future neutrino
observatories offered complementary detection techniques: a
megaton water Cherenkov detector such as MEMPHYS or
Hyper-Kamiokande, a liquid scintillator like LENA or a liquid
argon Time Projection Chamber (TPC) like GLACIER. See e.g.
\cite{Autiero:2007zj} for a detailed review on those large-scale
future detectors. The interest of such detectors, for a supernova
neutrino signal, has been showed e.g. in \cite{Fogli:2004ff} for
the water-Cherenkov type or in \cite{GilBotella:2003sz} for a
liquid Argon type. We focus on a detector of the latter kind since
it has the possibility to detect electron neutrino by
charged-current, through the reaction
\begin{eqnarray}
\nu_e + ^{40}{\rm Ar} \rightarrow ^{40}{\rm K^*} + e^-
\label{e:argon}
\end{eqnarray}
therefore yielding a very clear signal complementary of the
inverse $\beta$ of the water Cherenkov detectors.

To calculate the number of events received in such liquid Argon
detector (Icarus-like), we use the cross section given
in~\cite{GilBotella:2003sz} for the charged-current reaction
(\ref{e:argon}). For the detection of the $\mu-\tau$ resonance
effect in a liquid Argon type experiment, we calculate the usual
differential number of neutrinos detected at a distance D from the
supernova: \be \frac{d^2N_{events}}{dE_{\nu}dt}
=\frac{N_{targets}}{4\pi
D^2}\frac{d^2N_{\nu}}{dE_{\nu}dt}\sigma(E_\nu) \label{diffevent}
\ee where $N_{targets}$ is the number of targets in the detector,
D is the distance from the supernova, $E_{\nu}$ is the neutrino
energy, t is the detector time, $d^2N_{\nu}/dE_{\nu}dt$ the
differential flux of neutrinos arriving on Earth, for a given
detector time and a given neutrino energy, and $\sigma(E_\nu)$ is
the cross section of the reaction process considered. For a
galactic explosion, we suppose that the supernova explosion will
happen at the distance $D=10$kpc. Our calculations do not include
Earth matter effects \footnote{Because they depend upon the the
position of the supernova with respect to the detector when the
event occurs, we do not consider them in the neutrino signal in
the detector. A study of their effects is given for instance in
\cite{Dasgupta:2008my}.} and the Poisson error due to the finite
number of detected events.

For the signal really obtained in the detector, note that the
energy resolution reached in the detector is good enough to appreciate precisely
the $\mu-\tau$ resonance effect. Indeed, the energy of the
neutrino will not be directly obtained. It will be accessible via
the electron energy of the reaction of Eq.(\ref{e:argon}).

In a realistic supernova environment, after the post-bounce, both
the density profile and the luminosity are decreasing. As seen in
the previous section, this affects the interplay between the
$\mu-\tau$ resonance and the $\nu-\nu$ interaction. If the density
decreases, the $\mu-\tau$ resonance tends to get closer from the
nascent iron core with $r_{\mu \tau} \sim
\lambda_r^{\frac{1}{3}}$. When the luminosity decreases, the
bipolar region shrinks as $r_{bip} \sim \L_\nu^{\frac{1}{4}}$. Those
two phenomena are therefore opposite and tend to provoke a status
quo concerning the type of behaviour of the electron neutrino
flux. The way the $\nu_e$ flux energy distribution is modified
with time depends on the relative variation speed of the
parameters $\lambda_r$ and $\L_\nu$.

\begin{figure}[H]
\vspace{.6cm}
\centerline{\includegraphics[scale=0.3,angle=0]{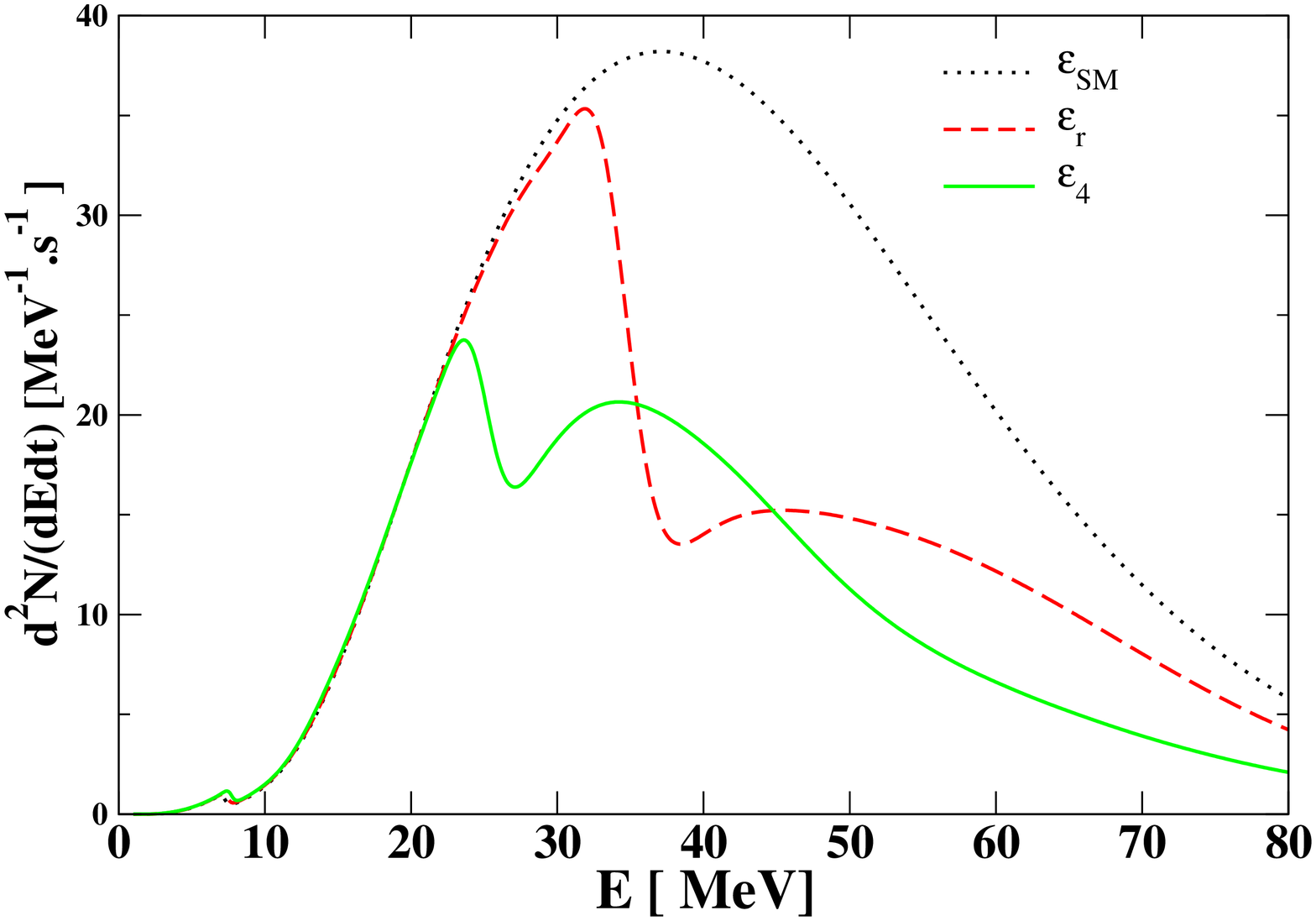}\hspace{.3cm}
\includegraphics[scale=0.3,angle=0]{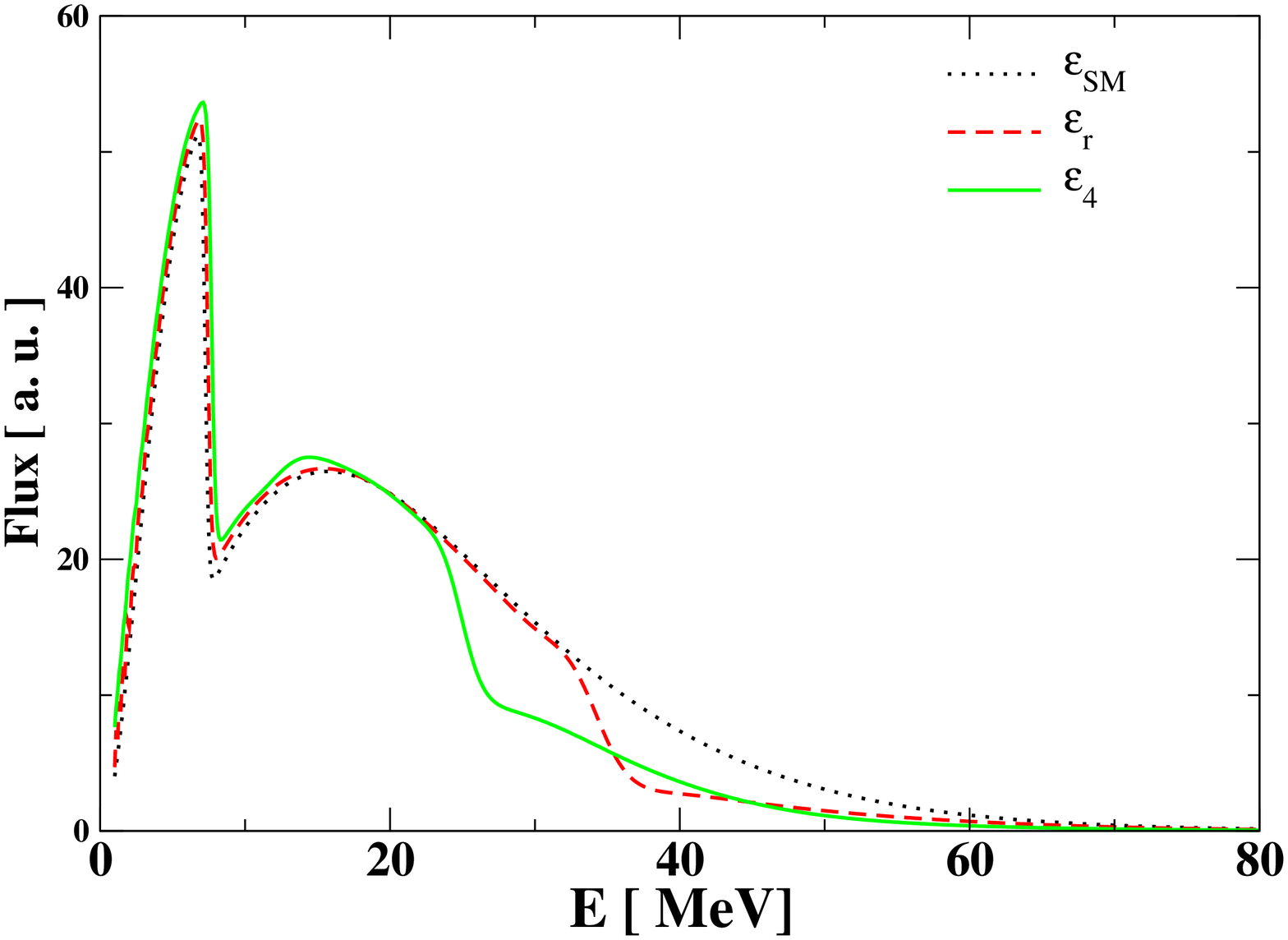}\hspace{.2cm}}
\caption{ In this case, the time after
post-bounce is supposed to be t=3s, the initial luminosity taken
is $L_{\nu}(t=3s) = 4.24 \times$ 10$^{51}$ erg.s$^{-1}$, the
density parameter is $\lambda_r$ = 1.14 $\times$ 10$^{6}$ and
$\theta_{23}$ = 40$^{\circ}$. Left figure: differential number of events on Earth in a
Liquid Argon detector of 70 kton size as a function of the energy
for three different $\varepsilon$. Right figure: corresponding $\nu_e$
flux on Earth as a function of the energy for three different
$\varepsilon$.} \label{fig:event3s}
\end{figure}

\begin{figure}[H]
\vspace{.6cm}
\centerline{\includegraphics[scale=0.3,angle=0]{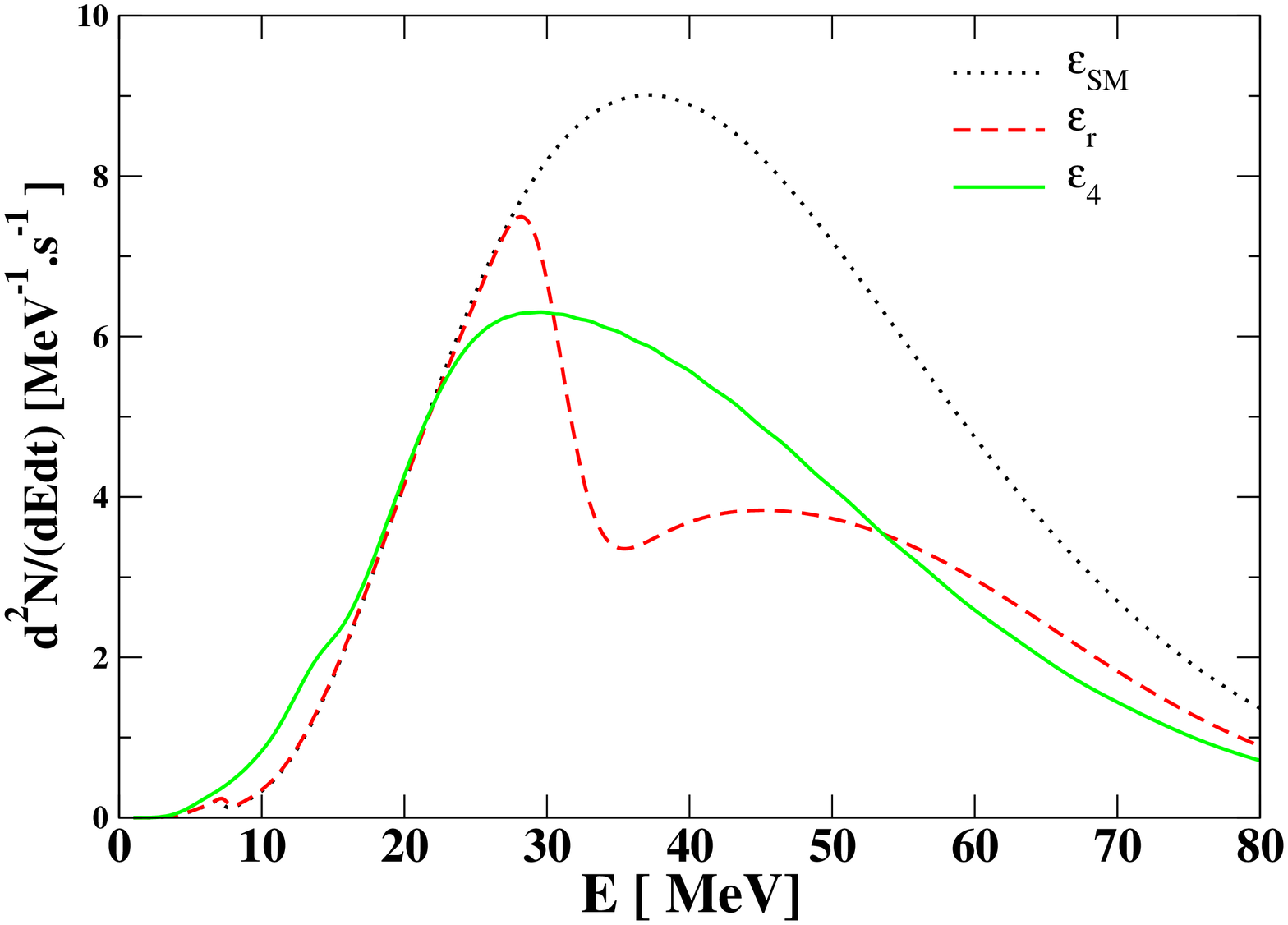}\hspace{.3cm}
\includegraphics[scale=0.3,angle=0]{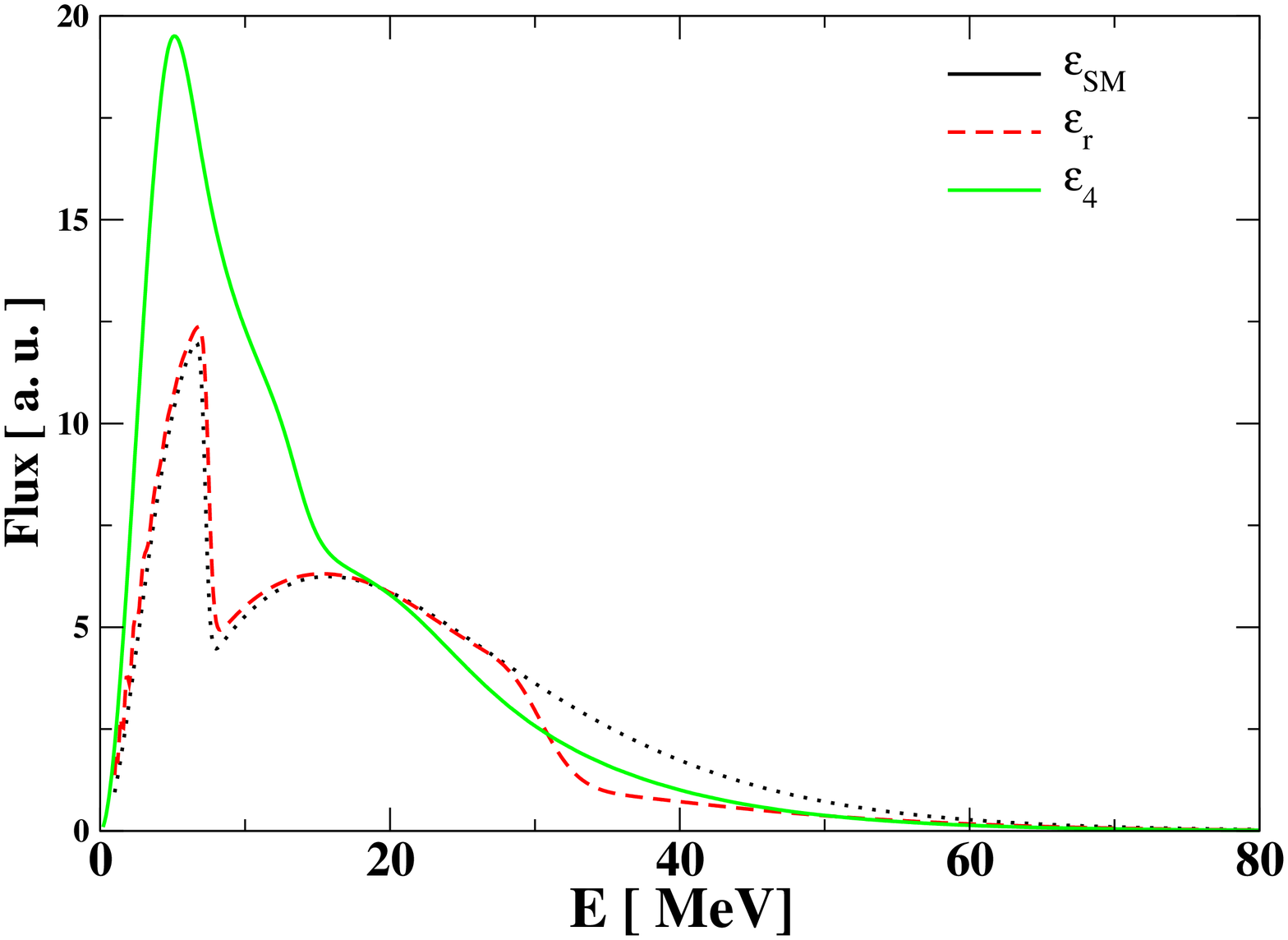}\hspace{.2cm}}
\caption{ In this case, the time after
post-bounce is supposed to be t=8s, initial luminosity taken is
$L_\nu$(t=8s) = 10$^{51}$ erg.s$^{-1}$, the density parameter is
$\lambda_r$ = 5.7 $\times$ 10$^{5}$ and $\theta_{23}$ =
40$^{\circ}$. Left figure: differential number of events on Earth in a
Liquid Argon detector of 70 kton size as a function of the energy
for three different $\varepsilon$.
Right figure: corresponding $\nu_e$ flux on Earth as a function of
the energy for three different $\varepsilon$.} \label{fig:event8s}
\end{figure}

Of course, the absolute value of the flux will undergo a global
exponential decrease with time. In the literature tackling the
question of dynamical density profile induced by shock-waves, the
difference of density during a time variation of about 5s, tends
to be of a factor $\sim 2$, for the part of the density which can
be well approximated by a power-law. See e.g.
\cite{Lunardini:2003eh} for the density profile as a function of
time. In Figs.(\ref{fig:event3s}) and (\ref{fig:event8s}), we show
the differential number of events received in a 70kton liquid
argon type detector and the associated flux for a post-bounce time
of 3 and 8 s. We study three distinct values for the radiative
correction term $\varepsilon$: the SM value
$\varepsilon_{SM}=5.4\times10^{-5}$ whose $N_{\mu \tau}$ will stay
in region I for these two times, an intermediate value
$\varepsilon_r = 10^{-3}$ which corresponds for the $\nu_e$ flux
to a case II behaviour until 8s, and the typical maximum value
obtained in the SUSY framework, i.e $\varepsilon_4=2\times10^{-2}$
which corresponds for the $\nu_e$ flux to case III behaviour
during most of the cooling phase.

For $t=3s$, the neutrino fluxes with $\varepsilon_r$ or
$\varepsilon_4$ entering the detector show important differences
with the SM case. While for $\varepsilon_r$ the difference with
$\varepsilon_{SM}$ appears for an energy of $\simeq 32$MeV as a
sizeable decrease of flux, the difference for $\varepsilon_4$
appears at a lower energy $\simeq 22$MeV. Looking at the differential number of events, we see
that the usual spectral split feature becomes almost invisible
except in the case of $\varepsilon_4$ where a very small bump remains.
On the contrary the effects of the $\mu-\tau$ resonance at higher
energies are much more visible. In the case of $\varepsilon_r$, a new spectral split feature appears. A type B behaviour starts at $\simeq 32$ MeV and leads to a type C behaviour from $\simeq 38$ MeV. For $\varepsilon_4$, the flux undergoes a diminution from $\simeq 22$MeV but this split feature is not as salient as the previous one. Nevertheless, we see that in the $\varepsilon_4$ case, the number of events is globally smaller than the $\varepsilon_r$ case. With respect to the SM case, the difference in the number of events corresponds in average to a factor $\sim$2 from $\simeq 22$ MeV.

For $t=8s$, the neutrino fluxes with $\varepsilon_r$ or
$\varepsilon_4$ display a modification w.r.t the SM
case similarly to time $t=3$s. Concerning the flux
with $\varepsilon_r$, it leaves the type A behaviour for a lower
energy equals to $\simeq 28$ MeV. this can be understood recalling Eq.(\ref{e:ratioradius}).
Indeed, since $\lambda_r$ has been multiplied by a factor 0.5 and
the luminosity has decreased by a factor $\simeq 4$, the relative
position of $r_{\mu \tau}$ over $r_{bip}$ has increased by a
factor $\simeq 4^{\frac{1}{4}} 2^{-\frac{1}{3}} > 1$.
Consequently, a lower energy is required for the type B behaviour
to start. Concerning the flux with $\varepsilon_4$,
the relative difference w.r.t. the flux in the SM case is more
important at low energies. Nevertheless, the luminosity being exponentially
decreasing with time and the low value of the cross-section at low energies will prevent to see the effects of $\varepsilon_4$ on the number of events for those energies. Finally, we can see on the number of events figure, that the $\varepsilon_r$ curve presents globally a more pronounced difference w.r.t to the SM case than the $\varepsilon_4$ curve.

\begin{figure}[H]
\vspace{.6cm}
\centerline{\includegraphics[scale=0.3,angle=0]{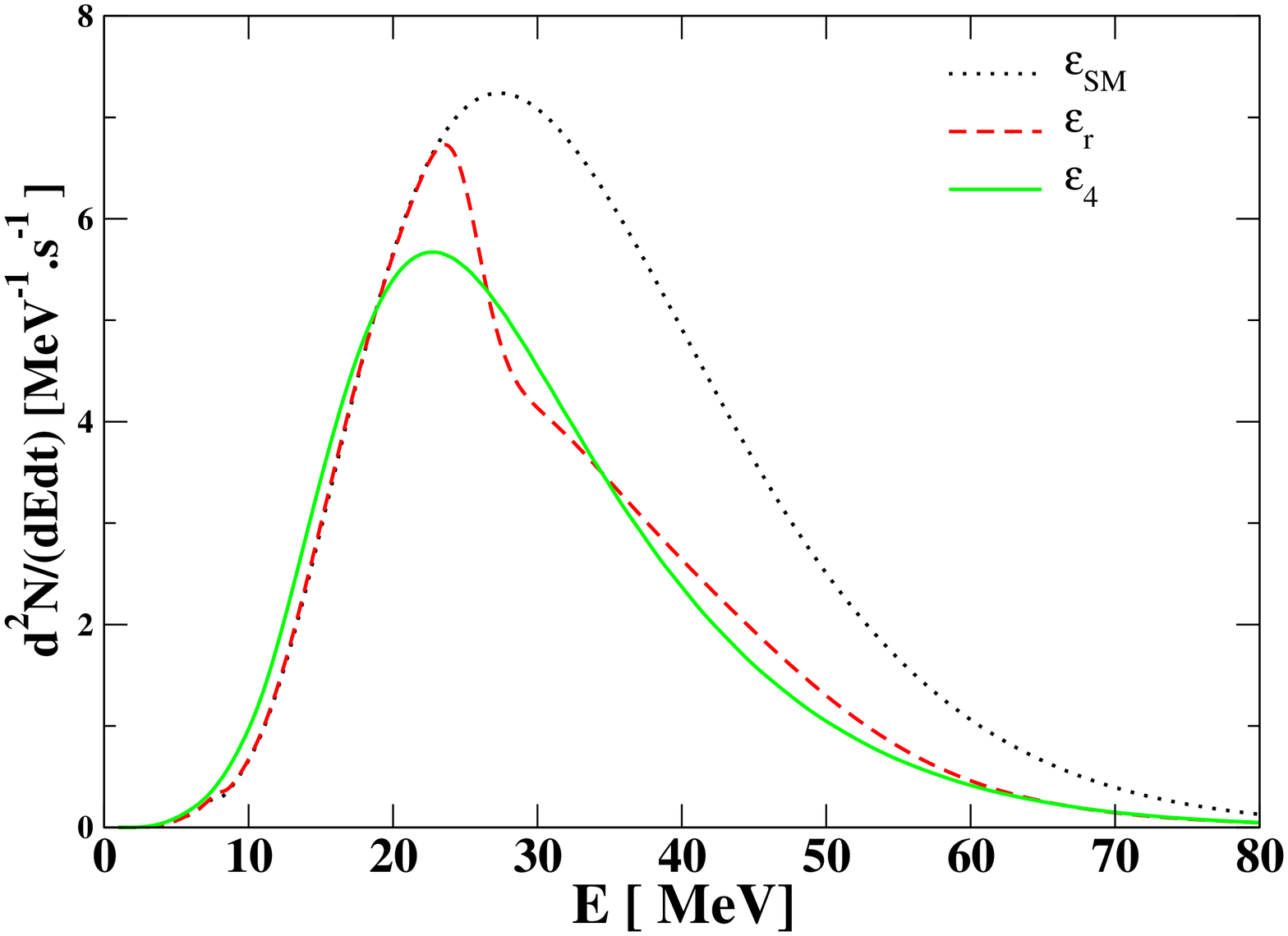}\hspace{.3cm}
\includegraphics[scale=0.3,angle=0]{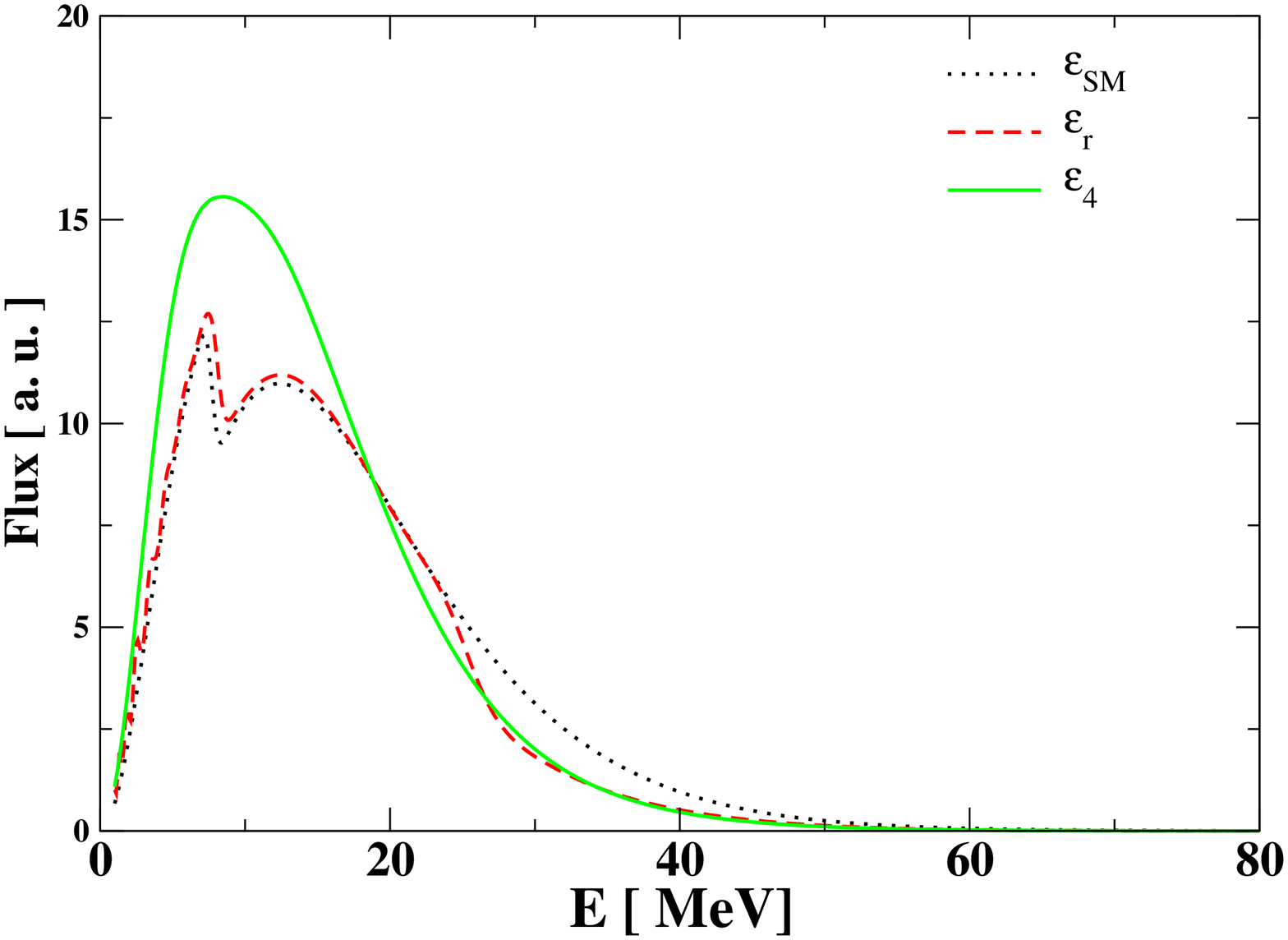}\hspace{.2cm}}
\caption{ In this case, the time after
post-bounce is supposed to be t=8s, initial luminosity taken is
$L_{\nu}$(t=8s) = 10$^{51}$ erg.s$^{-1}$, the density parameter is
$\lambda_r$ = 5.7 $\times$ 10$^{5}$ and $\theta_{23}$ =
40$^{\circ}$. In these figures the hierarchy in temperature is $ \langle E_{\nu_e}\rangle =12$ MeV, $\langle E_{\bar{\nu}_e} \rangle =15$ MeV, $\langle E_{\nu_{x}} \rangle =18$ MeV. Left figure: differential number of events on Earth in a
Liquid Argon detector of 70 kton size as a function of the energy
for three different $\varepsilon$. Right figure: corresponding $\nu_e$ flux on Earth as a
function of the energy for three different $\varepsilon$.}
\label{fig:event8s-hierarchy}
\end{figure}

In Fig.(\ref{fig:event8s-hierarchy}), we take other values for the energy hierarchy in agreement with the numerical simulations performed in \cite{Keil:2002in}: i.e. $ \langle E_{\nu_e}\rangle =12$ MeV, $\langle E_{\bar{\nu}_e} \rangle =15$ MeV, $\langle E_{\nu_{x}} \rangle =18$ MeV respectively. Concerning the flux with $\varepsilon_r$, the difference w.r.t the SM case appears from $\simeq 23$ MeV. For $\varepsilon_4$, the spectral split feature has vanished and the flux presents a much higher value w.r.t the SM case around 10 MeV. Concerning the number of events, both $\varepsilon_r$ and $\varepsilon_4$ curves move away from the SM one from around 20 MeV. We therefore conclude that even with this hierarchy of temperatures, sizeable effects are present on the differential number of events received in a Liquid Argon detector.

Though we did not look at the time evolution $\nu_e$ flux for
given energies during the whole cooling phase, it seems that the
global pattern of the fluxes will not be much modified with time,
contrary to what may happen for the electron anti-neutrino flux
\cite{Gava:2009pj}. In this case, to distinguish w.r.t. the SM
case would require the identification of a particular energy
distribution of the flux.

Considering the $\nu_e$ flux, for the $\varepsilon_r$ case, the
spectral split feature seems peculiar enough to be recognized: at
40 MeV, the ratio between the SM flux and the $\varepsilon_r$ flux
is more or less constant. One could also look at the time
evolution of the flux for an energy of 30 MeV and 40 MeV. In the
SM case, the flux at 40 MeV will be systematically higher than the
one at 30 MeV. In the case of a large $\varepsilon$, the flux at
40 MeV will be systematically lower. For the $\varepsilon_4$ case,
the low energy bump may be visible for early time because of an
important global luminosity. Therefore, the presence of a spectral
split at high energy in the differential number of events may be a
signature of a large radiative correction term and therefore a
beyond standard physics.

Using the signal in the electron anti-neutrino channel with a
Cherenkov detector, one would be able to obtain precise
information on the density profile as a function of time. In
addition, with enough precision on the neutrino parameters, like
the octant of $\theta_{23}$, as well as Supernova parameters
partly given by the other channels (electron anti-neutrino via
inverse $\beta$ reaction, all flavour flux via neutral current),
this could ideally lead to new constraints for the SUSY parameters
via the analytical calculation of the radiative correction as in
\cite{Gava:2009gt,Roulet:1995ef}. Equivalently, the absence of
such characteristic imprints would also lead to constraints if no
other effects like turbulence modify the $\nu_e$ flux existing
inside the supernova as will be discussed in the next section.

Finally, note that the impact of the interplay between $V_{\mu \tau}$ and the collective effects has been observed
 on the Diffuse Supernova Neutrino Background (DSNB) \cite{Galais:2009wi}. Such interplay tends to provoke important effects even in the Standard Model thanks to the redshift.
\section{Discussions}
\label{sec:discussion}

In this section we discuss on how much these results can be seen as realistic.

We have performed a calculation of the neutrino propagation in a
core-collapse supernova including the neutrino-neutrino
interaction and a large mu-tau index of refraction coming from
SUSY radiative corrections. Such combination in certain conditions
of matter density, energy, luminosity and oscillation parameters
implies a resonance, associated to the loop correction potential.
This occurs in the same region as the bipolar resonance.
Consequently, the $\nu_\mu$ and $\nu_\tau$ oscillation
probabilities are modified and the low-resonance becomes much more
adiabatic leading to an important inversion of fluxes, visible on
the electron neutrino flux on Earth as a function of energy. Such
characteristic imprints of a large $V_{\mu \tau}$, i.e of a BSM
physics, would be also sizeable on the number of events in a
Liquid Argon detector type.

In a supernova environment, such interplay has been first studied
for a given energy in \cite{EstebanPretel:2007yq}. In order to
obtain a $\mu-\tau$ resonance with a radiative correction term in
the SM framework, the authors considered a very high density
profile. This way, such resonance happens well outside the bipolar
transition zone and a clear, easily understandable, behaviour is
exhibited. Unfortunately, it has been shown
in~\cite{EstebanPretel:2008ni}, for the multi-angle and, therefore,
general case, that such high density prevents collective effect from
taking place. Consequently, we have to check if the density we use is low enough for the multiangle matter effect to be neglected. In \cite{EstebanPretel:2009is}, they found that the condition:
\be
Y_e(r_{syn})\lambda_r \frac{R}{r_{syn}} < \mu_0
\ee
must be fulfilled. Using Eq.(11) of \cite{EstebanPretel:2007ec}, we find in our case, $\mu_0\simeq 2.6\times 10^5$ km$^{-1}$. Therefore, having $Y_e=0.5$ and assuming $r_{syn} \simeq 100$ km the condition becomes : $\lambda_r < 5.2\times 10^6$ km$^{-1}$.
Since $\lambda_r=5.7 \times 10^5$ km$^{-1}$, we can indeed safely neglect the matter effects.

The advantage of having a large $V_{\mu \tau}$ potential due to a
large SUSY radiative correction is significant for multiple
reason. In this case, a very high density is not required for the
$\mu-\tau$ resonance to be outside the zone where collective
effects predominate. Therefore, we naturally avoid the problem of the high
density effect cited above.

Concerning our density profile, we use an analytical power law
which does not take into account shock-waves effects. Nevertheless
we believe such dynamical effects are not determinant in our case.
Indeed, we checked \cite{Jim:2009ab} that the shock-wave effects
such as depicted in \cite{Gava:2009pj} do not affect the $\nu_e$
flux\footnote{The results found in \cite{Galais:2009wi} show explicitly that the presence of the shockwave does not
 prevent the consequences on the electron neutrino flux of the interplay between $V_{\mu\tau}$ and the collective effects.}.
In inverted hierarchy, $\nu_e$s do not encounter the H-resonance,
and therefore do not undergo the effect of adiabaticity breaking
and/or multiple resonance which tends to smear out the neutrino
flux distribution. Moreover, looking at the $\nu_e$ flux during
the cooling phase not too early, allows the shock wave to affect
the density profile well after 1000 km, i.e well after that the
$\mu-\tau$ resonance and the $\nu-\nu$ interaction occur.
Similarly, the L resonance is not supposed to be significantly
affected by the shock-wave effects \cite{Kneller:2007kg}.

Concerning the collective effects, even if the matter density is
not so high, multi-angle should be prevented of triggering
decoherence because of the asymmetry between the $\nu_e$ and
$\bar{\nu}_e$ flux cause by deleptonization
\cite{EstebanPretel:2007ec}. Consequently, single-angle behaviour
may well be typical for realistic SN conditions as long as this
flux asymmetry remains.

Note that we have checked that the influence of $\delta$ is
negligible in this particular case where we focus on the electron
neutrino signal in the inverted hierarchy. Moreover, we took equal
luminosities for $\nu_\mu$ and $\nu_\tau$ flux
\cite{Balantekin:2007es}. As long as it is non zero, the influence
of $\theta_{13}$ on the $\nu_e$ flux in inverted hierarchy is very
small, possibly below the experimental uncertainties.

In our paper, we also made the assumption of a constant electron
fraction. This approximation is justified when we restrict
ourselves to a maximum value of $\simeq 2\times 10^{-2}$ for
radiative term $\varepsilon$. In this case, no non-standard
interaction (NSI) resonance will occur as shown
in~\cite{EstebanPretel:2009is}.~\footnote{Note that our value of
$\varepsilon$ doesn't match exactly to the value of the NSI
diagonal term $\varepsilon_{\tau\tau}$ given
in~\cite{EstebanPretel:2009is}.}

Similar interplay should be in principle present in the
anti-neutrino channel but as pointed out in \cite{Gava:2009pj},
the presence of shock-waves effects tend to render the H-resonance
non adiabatic and eventually create a multiple resonance effect
which can smear out the electron anti-neutrino signal. The
addition of such important radiative correction matter potential
should disturb the signal but full numerical calculation as first
performed in should be done to witness the impact on the electron
anti-neutrino flux.

Concerning the normal hierarchy, the effects of the $\mu-\tau$
resonance tend to be much smaller as no bipolar transition occurs.
Nevertheless, the case where spectral splits take place  in this
hierarchy \cite{Dasgupta:2008cd,Dasgupta:2009mg} should be
investigated in the presence of a large $\varepsilon$.

Finally, we discuss the most important effect that could prevent
the salient modification of the $\nu_e$ flux induced by the
$\mu-\tau$ resonance and $\nu-\nu$ interplay. A realistic density
profile should present possible stochastic matter fluctuations due
to hydrodynamical instabilities created in the wake of a shock
front. This has been studied in several works
\cite{Fogli:2006xy,Friedland:2006ta} but the precise size of such
fluctuations remains an important open issue. In principle, the
$\mu-\tau$ resonance should not be affected. The question whether
the L-resonance will be affected or not, depends on the
fluctuation scale i.e if the damping effects are significant on
the flavour transition pattern. Associated with this issue, taking
into account a non-spherical symmetry for the neutrino emission
\cite{Dasgupta:2008cu} as well with matter density profile
\cite{Kneller:2007kg}, may be relevant, for example, in the
presence of hydrodynamic turbulence in the background of ordinary
matter. The framework considered in this paper, defined in
Sec.~\ref{sec:theoframe}, allows us to think that our results
should contain the main features of a realistic signal.

\section{Conclusion}
\label{sec:conclusion}

In this paper, we worked in the SUSY framework. If new phenomena
compatible with this framework are discovered on the supernova
neutrino fluxes, one of the major challenges will be to find out
the underlying model and to measure its parameters as precisely as
possible. This can be done for example by using codes in which we
consider the phenomenological constraints~\cite{nmssmtools}. It
will be necessary to investigate the
precision\footnote{Considering two-loop corrections would give the
size of the approximation made.} with which SUSY model
parameters~\cite{Gava:2009gt} can be derived from these
measurements. Finally, if Supersymmetry is realised in Nature, it
should be possible to constrain the SUSY parameters thanks to the
discovery of sparticles at TeV colliders. For a galactic
explosion, using a fully detailed numerical simulation, supernova
neutrino fluxes could ideally lead to constraints on the SUSY
parameters in the inverted hierarchy case. In the future, the
results presented in this paper should be applied to a larger
variety of models and confronted to some other data from the Large
Hadron Collider (LHC) and the International Linear Collider (ILC).
Finally, let us say that the $\mu-\tau$ resonance effect on the
energy distribution of the electron neutrino flux could impact on
the number of events for relic supernova neutrinos as in
\cite{Galais:2009wi}. In this case, it could give an even lower
total red-shift-integrated number of events than in the SM case
which can be consistent with observations. Moreover, a large
$\varepsilon$ could also impact the heavy elements nucleosynthesis
in supernova, and more precisely on the electron fraction through
the r-process. Supernovae can definitively be seen more than ever,
as an open window towards beyond Standard Model physics.

\section*{Acknowledgements}

The authors are grateful to J. Kneller and S. Galais for numerical
checkings, discussions and careful reading of the manuscript. J.
Gava thanks also C. Volpe for useful discussion.

\textit{}

\end{document}